\documentclass[11pt,a4paper]{article}
\usepackage{jheppub}
\usepackage[utf8]{inputenc}
\usepackage{latexsym, graphicx} 
\usepackage{amsmath,amsfonts,amssymb}
\usepackage{mathrsfs}
\usepackage{tensor}
\usepackage{cancel}

\newcommand{\Tr}{\text{Tr}}
\newcommand{\reg}{\text{reg}}
\newcommand{\ct}{\text{ct}}
\newcommand{\ren}{\text{ren}}
\newcommand{\CS}{\text{CS}}

\newcommand*{\Jt}{\mathtt{J}}
\newcommand*{\bJt}{\bar{\mathtt{J}}}

\newcommand*{\sa}{\sf{a}}
\renewcommand*{\sb}{\sf{b}}
\renewcommand*{\sc}{\sf{c}}
\newcommand*{\sA}{\sf{A}}
\newcommand*{\sB}{\sf{B}}
\newcommand*{\sC}{\sf{C}}

\newcommand*{\ea}{e^{\sA}}
\newcommand*{\eb}{e^{\sB}}
\newcommand*{\ec}{e^{\sC}}
\newcommand*{\oa}{\omega^{\sA}}

\newcommand*{\bA}{\bar{A}}
\newcommand*{\ba}{\bar{a}}
\newcommand*{\bG}{\bar{G}}

\newcommand*{\bh}{\bar{h}}
\newcommand*{\bF}{\bar{F}}
\newcommand*{\blambda}{\bar{\lambda}}
\newcommand*{\bPsi}{\bar{\Psi}}
\newcommand*{\bL}{\bar{L}}
\newcommand*{\bz}{\bar{z}}
\newcommand*{\bw}{\bar{w}}
\newcommand*{\te}{\tilde{e}}
\newcommand*{\tea}{\tilde{e}^{\sf{a}}}
\newcommand*{\tomega}{\tilde{\omega}}

\title{Holographic boundary actions in AdS$_3$/CFT$_2$ revisited}
\author[]{Kevin Nguyen}
\emailAdd{kevin.nguyen@kcl.ac.uk}
\affiliation[]{Department of Mathematics, King's College London, London, United Kingdom}

\abstract{
The generating functional of stress tensor correlation functions in two-dimensional conformal field theory is the nonlocal Polyakov action, or equivalently, the Liouville or Alekseev--Shatashvili action. I review its holographic derivation within the AdS$_3$/CFT$_2$ correspondence, both in metric and Chern--Simons formulations. I also provide a detailed comparison with the well-known Hamiltonian reduction of three-dimensional gravity to a flat Liouville theory, and conclude that the two results are unrelated. In particular, the flat Liouville action is still off-shell with respect to bulk equations of motion, and simply vanishes in case the latter are imposed. The present study also suggests an interesting re-interpretation of the computation of black hole spectral statistics recently performed by Cotler and Jensen as that of an explicit averaging of the partition function over the boundary source geometry, thereby providing potential justification for its agreement with the predictions of a random matrix ensemble.}

\begin{document}

\maketitle
\flushbottom

\section{Introduction}
Boundary actions derived in the context of three-dimensional gravity with anti-de Sitter (AdS) asymptotics have a relatively long history\footnote{Connections between SL(2,$\mathbb{R}$) Chern-Simons theory and various boundary actions discussed in the present paper also appeared in earlier work of Verlinde \cite{Verlinde:1989ua}, often viewed as a precursor of the AdS/CFT correspondence.}, starting with the classic work of Coussaert, Henneaux and van Driel \cite{Coussaert:1995zp} a few years before the advent of the AdS/CFT correspondence \cite{Maldacena:1997re,Gubser:1998bc,Witten:1998qj}. Roughly speaking, these authors reduced the classical dynamics of pure three-dimensional gravity in its Chern--Simons (CS) formulation to that of a Liouville theory, one of the simplest non-rational two-dimensional conformal field theory (CFT). This raised the hope that quantum Liouville theory would describe at least part of three-dimensional quantum gravity and more specifically, that it would yield a description of the microstates underlying the Ba\~nados--Teitelboim--Zanelli (BTZ) black holes and the associated Bekenstein--Hawking entropy \cite{Carlip:2005zn}. Several variations of this reduction have followed, either generalizing it or differing in the choice of boundary conditions \cite{Henneaux:1999ib,Rooman:1999km,Rooman:2000zi,Barnich:2013yka,Cotler:2018zff,Henneaux:2019sjx}.

The AdS/CFT correspondence is another approach to quantum gravity with AdS asymptotics which has met numerous successes and is by now a well-established subject. In essence, it supports the existence of a \textit{holographic} conformal field theory at the AdS conformal boundary, which encodes the information of the gravitational theory at a quantum level. The precise statement of the correspondence is an equality between the generating functional of the holographic CFT and the path integral of the gravitational theory \cite{Gubser:1998bc,Witten:1998qj},
\begin{equation}
\label{GKPW}
Z_{\text{CFT}}\left[\tilde g_{ij},J\right]\equiv \int_{\substack{G_{\mu\nu}(\infty)=\tilde g_{ij}\\ \Phi(\infty)=J}} \mathcal{D}G_{\mu\nu}\, \mathcal{D}\Phi\, e^{-S_{\text{grav}}\left[G_{\mu\nu},\Phi\right]}\,.
\end{equation}
Here, $G_{\mu\nu}$ is the dynamical bulk metric while $\Phi$ refers to all other bulk fields arising in a given realization of the correspondence. Their boundary values $\tilde g_{ij}, J$ are fixed at the conformal boundary of spacetime and are further identified with the sources of the dual CFT. In particular, $\tilde g_{ij}$ sources the CFT stress tensor. However, the equality \eqref{GKPW} is at most formal since there is no available nonperturbative definition of its right-hand side, such that the holographic CFT is often said to \textit{define} nonperturbative quantum gravity. In the case of three bulk dimensions, the central charge of the holographic CFT directly relates to Newton's constant and the AdS curvature length~$\ell$ via the Brown--Henneaux formula \cite{Brown:1986nw},
\begin{equation}
	\label{c}
	c=\frac{3\ell}{2 G_N}\,.
\end{equation}
Therefore, the limit of large central charge corresponds to the classical limit in the gravitational theory, in which case the gravitational path integral reduces to a sum over classical saddle points,
\begin{equation}
\label{GKPW classical}
\lim\limits_{c \to \infty} Z_{\text{CFT}}\left[\tilde g_{ij}, J\right] \approx \sum_{n} e^{-S_{\text{grav}}[G^{(n)}_{\mu\nu},\Phi^{(n)}]}\,,
\end{equation}
where each field configuration $\lbrace G^{(n)}_{\mu\nu}, \Phi^{(n)}\rbrace$ is a classical saddle of $S_{\text{grav}}$ that satisfy the prescribed boundary conditions. I will refer to \eqref{GKPW classical} as the Gubser--Klebanov-Polyakov--Witten (GKPW) dictionary \cite{Gubser:1998bc,Witten:1998qj}. In concrete realizations of the AdS/CFT correspondence derived within string theory, the action $S_{\text{grav}}$ describing the classical regime of the theory is that of a supergravity or dimensional reduction thereof. Of particular interest is the case where all sources except $\tilde g_{ij}$ are turned off, i.e., when $J=0$. With these boundary conditions and working in the classical limit \eqref{GKPW classical}, it is consistent to set to zero all bulk fields $\Phi$ such that only pure gravity remains. In that case, the left-hand side of \eqref{GKPW classical} reduces to the generating functional of stress tensor correlation functions at large central charge. Two-dimensional conformal symmetry is so restrictive that the latter can only depend on the value of the central charge but is otherwise universal. Explicitly, it is given by the nonlocal Polyakov action \cite{Polyakov:1981rd}, which can alternatively be written as the action of a Liouville theory. More or less successful derivations of the Polyakov action within AdS$_3$/CFT$_2$ have been given in \cite{Skenderis:1999nb,Krasnov:2000zq,Krasnov:2001cu,Manvelyan:2001pv,Banados:2002ey,Banados:2004nr,Carlip:2005tz}.

The primary goal of the present work is to review the emergence of the Polyakov and Liouville actions within AdS$_3$/CFT$_2$ by direct application of the GKPW dictionary \eqref{GKPW classical}, and to compare it with the Hamiltonian reduction of three-dimensional gravity to flat Liouville theory in the tradition initiated by Coussaert, Henneaux and van Driel \cite{Coussaert:1995zp}. The result of this comparative study is that these results are unrelated. In particular, the flat Liouville action is off-shell with respect to bulk equations of motion, and vanishes if the latter are imposed.

Another motivation for the present study is the recent computation due to Cotler and Jensen of the spectral statistics of BTZ black holes that found agreement with the predictions of a random matrix ensemble \cite{Cotler:2020ugk}. This raised a conceptual puzzle, since AdS$_3$ gravity is usually understood as being dual to a single CFT rather than some ensemble of theories \cite{Aharony:1999ti}. The starting point for their computation is a variant of the Hamiltonian reduction of AdS$_3$ gravity to (copies of) the Alekseev--Shatashvili action at the conformal boundary \cite{Cotler:2018zff}. The BTZ spectral statistics follow from the evaluation of a path integral weighted by this boundary action. I will explain how the Alekseev--Shatashvili action can alternatively be obtained from the classical GKPW dictionary \eqref{GKPW classical}, and show that in that context it describes curved boundary metrics acting as sources for the CFT stress tensor. This suggests a  possible re-interpretation of the path integral computation yielding the BTZ spectral statistics as that of an explicit averaging over the boundary geometry, or equivalently, as a computation in two-dimensional quantum gravity at the conformal boundary. This potentially resolves the aforementioned conceptual puzzle.

The paper is organized as follows. In order to guide our expectations, I review in section~\ref{section:Polyakov} the role played by the nonlocal Polyakov action as the universal generating functional of stress tensor correlation functions in any two-dimensional CFT. I also discuss its relation with the Liouville and Alekseev--Shatashvili actions. In section~\ref{section:metric}, I review its holographic derivation within the AdS$_3$/CFT$_2$ correspondence, originally given by Carlip in the metric formalism \cite{Carlip:2005tz}. In section~\ref{section:Banados}, I provide the analogue derivation in the Chern--Simons formulation of three-dimensional gravity, thereby bringing to completion the work of Ba\~nados, Chandia and Ritz \cite{Banados:2002ey}. I compare it with the Hamiltonian reduction of Coussaert, Henneaux and van Driel \cite{Coussaert:1995zp} in section~\ref{section:Hamiltonian reduction} and conclude that these approaches are technically and physically inequivalent. In section~\ref{section:Cotler}, I discuss a variant Hamiltonian reduction due to Cotler and Jensen and explain how the resulting boundary action can alternatively be obtained from the GKPW dictionary \eqref{GKPW classical}. I end with a discussion of the results and a few open questions. In particular, I suggest a possible re-interpretation of the recent agreement found between black hole spectral statistics and ensemble averaging \cite{Cotler:2020ugk}. 

\paragraph{Conventions.} 
Conformal field theory is discussed in euclidean signature with complex coordinates $z,\bz$ and integral measure $d^2z=dz \wedge d\bz/2i$. This allows straightforward analytic continuation to Lorentzian signature without modification of the formulas. I also use the shorthand notations $T\equiv -2 \pi T_{zz}$ for the holomorphic component of the stress tensor, and $\partial\equiv \partial_z\,, \bar \partial\equiv \partial_{\bz}$ for simplicity. In the gravitational setup, greek indices $\mu, \nu,...$ refer to three-dimensional spacetime coordinates, while latin indices $i,j,...$ refer to two-dimensional coordinates at the conformal boundary. Similarly, uppercase letters $\sA, \sB, ...$ and lowercase letters $\sa, \sb,...$ refer to the corresponding three- and two-dimensional tangent spaces. 

\section{The Polyakov action and its many forms}
\label{section:Polyakov}
In any euclidean two-dimensional conformal field theory, the generating functional $W[g_{ij}]$ of \textit{connected} stress tensor correlation functions is defined from the generating functional $Z[g_{ij}]$ via the standard relation
\begin{equation}
Z[g_{ij}]=e^{-W[g_{ij}]}\,,
\end{equation}
where the two-dimensional background metric $g_{ij}$ plays the role of a source for the stress tensor $T_{ij}$. Functional differentiation of $W[g_{ij}]$ yields the connected correlators,
\begin{equation}
\label{eq:T correlators}
\langle T_{ij}(x_1) ... T_{mn}(x_n) \rangle_{g_0} =-\frac{(-2)^{n}}{\sqrt{g(x_1)} ... \sqrt{g(x_n)}}\, \frac{\delta^n W}{\delta g^{ij}(x_1) ... \delta g^{mn}(x_n)}\Big|_{g=g_0}+...\,,
\end{equation}
where the dots refer to contact terms resulting from functional differentiation of the metric determinant of the type
\begin{equation}
\frac{\delta}{\delta g^{ij}(x_k)}\left( \frac{1}{\sqrt{g(x_l)}}\right)\,.
\end{equation}

\paragraph{General covariance.} Stress tensor correlators are fully constrained by conformal symmetry, the only dependence on a given theory occurring through the central charge $c$ \cite{Belavin:1984vu}. Up to its value, the generating functional $W\left[g_{ij}\right]$ is therefore universal. Polyakov's starting point for its construction is the anomalous trace of the stress tensor expectation value on a space with arbitrary background metric $g_{ij}$ and curvature $R$ \cite{Polyakov:1981rd},
\begin{equation}
\frac{c}{24\pi} R=g^{ij} \langle T_{ij} \rangle=\frac{2}{\sqrt{g}}\,g^{ij} \frac{\delta W}{\delta g^{ij}},
\end{equation}
where the last equality follows from the definition \eqref{eq:T correlators}. Integrating this equation while insisting on keeping manifest \textit{general covariance}, Polyakov obtained \cite{Polyakov:1981rd}
\begin{subequations}
	\label{eq:Polyakov action}
	\begin{align}
		W\left[g_{ij}\right]&=\frac{c}{96\pi} \int d^2x\, d^2y\, \sqrt{g(x)}\, \sqrt{g(y)}\, R(x) G(x,y) R(y)\\
		&=\frac{c}{96\pi} \int d^2x\, \sqrt{g(x)}\, R(x) \frac{1}{\square} R(x),
	\end{align} 
\end{subequations}
where $G(x,y)$ is the Green function solution to
\begin{equation}
	\square G(x,y)=\frac{\delta^{(2)}(x-y)}{\sqrt{g(x)}}.
\end{equation}
The Polyakov action \eqref{eq:Polyakov action} is manifestly nonlocal in the background metric $g_{ij}$. It can be put in an alternative form through the introduction of an auxiliary variable $\phi$ solving
\begin{equation}
	\label{eq:boxphi}
	\square \phi=R,
\end{equation} 
such that the generating functional coincides with the action of a Liouville theory,
\begin{equation}
	\label{eq:Liouville action}
	W\left[g_{ij}\right]=\frac{c}{48\pi} \int d^2x\, \sqrt{g(x)}\, \left(\frac{1}{2}(\partial \phi)^2+\phi R\right)\,.
\end{equation} 
We stress that the Liouville field $\phi$ is not an independent variable, but rather a nonlocal functional of the metric through \eqref{eq:boxphi}. Said differently, the Liouville action \eqref{eq:Liouville action} reduces to the Polyakov action \eqref{eq:Polyakov action} only onshell, i.e., when the Liouville field satisfies its own equation of motion \eqref{eq:boxphi}. The stress tensor expectation value may be computed from \eqref{eq:Liouville action} by functional differentiation, and is found to coincide with the classical Liouville stress tensor 
\begin{equation}
	\label{eq:Tij}
	\langle T_{ij} \rangle=\frac{2}{\sqrt{g}} \frac{\delta W}{\delta g^{ij}}=\frac{c}{24\pi} \left[\frac{1}{2}\partial_i \phi\, \partial_j \phi-\nabla_i \nabla_j \phi+g_{ij} \left(\square \phi-\frac{1}{4} (\partial \phi)^2 \right)  \right]=T_{ij}^{\phi}\,.
\end{equation}
Consistently, one recovers the trace anomaly which we started from,
\begin{equation}
	\label{eq:trace anomaly}
	g^{ij} \langle T_{ij} \rangle =\frac{c}{24\pi}\, R\,.
\end{equation}
It is worth mentioning that covariance of \eqref{eq:boxphi} under a Weyl rescaling 
\begin{equation}
	\label{eq:Weyl rescaling}
	g_{ij} \mapsto e^{\omega} g_{ij}\,,
\end{equation}
implies that $\phi$ must transform by a shift $\phi \mapsto \phi- \omega$. It is therefore natural to interpret $\phi$ as the pseudo-Goldstone mode associated to broken Weyl symmetry. The expectation value \eqref{eq:trace anomaly}, or equivalently the configuration $\phi$ determined through \eqref{eq:boxphi}, labels one of the broken vacua. Due to explicit breaking of Weyl symmetry by the central charge $c$, this pseudo-Goldstone mode acquires a nonzero action \eqref{eq:Liouville action}. One can also compute higher-point correlations by application of \eqref{eq:T correlators}. For the two-point function on the euclidean plane, a straightforward computation yields
\begin{equation}
\langle T_{ij}(x) T_{mn}(y) \rangle_{\text{plane}}=-\frac{c}{48\pi^2} \left(\delta_{ij}\square^x -\nabla_i^x \nabla_j^x \right) \left(\delta_{mn} \square^y -\nabla_m^y \nabla_n^y \right) \ln \mu^2|x-y|^2,
\end{equation}
where $\mu$ is an arbitrary energy scale introduced such that the argument of the logarithm is dimensionless. Choosing complex coordinates
\begin{equation}
\label{eq:flat metric}
ds^2=dz\, d\bz,
\end{equation} 
one recovers in particular the standard expression 
\begin{equation}
	\langle T(z,\bz) T(w,\bar{w}) \rangle= \frac{c}{2(z-w)^4}\,.
\end{equation}

\paragraph{Holomorphic factorization.} The Polyakov action preserves general covariance which is however not one of the defining features of two-dimensional conformal field theories. Rather, one usually insists on holomorphic factorization as one of its basic principles \cite{Belavin:1984vu}. It turns out that holomorphic factorization and diffeomorphism invariance are actually incompatible. One can however pass from one formulation to the other by the addition of local counterterms to the generating functional $W[g_{ij}]$. Let's see how this works in practice. We consider a general expression for the background metric in some coordinates system~$(z,\bz)$,
\begin{equation}
\label{eq:metric z zbar}
ds^2= e^{\varphi(z,\bz)}\left(dz + \mu(z,\bz)\, d\bz\right)\left(d\bz+ \bar \mu(z,\bz)\, dz\right)\equiv e^\varphi\, d\hat{s}^2\,,
\end{equation}
where $\varphi, \mu, \bar \mu$ are arbitrary functions. With this form of the metric, it can be shown that the Polyakov action \eqref{eq:Polyakov action} admits the decomposition \cite{Verlinde:1989ua,Verlinde:1989hv,Knecht:1990wb}
\begin{equation}
W\left[g_{ij}\right]=K\left[\varphi,\mu,\bar \mu\right]+W\left[\mu\right]+\bar W\left[\bar \mu\right]\,.
\end{equation}
The first term $K\left[\varphi,\mu,\bar \mu\right]$ is known as the Quillen--Belavin--Knizhnik anomaly \cite{Quillen1985DeterminantsOC,Belavin:1986cy} and is responsible for the non-factorization of stress tensor correlators in generic backgrounds. Its explicit expression is 
\begin{equation}
	K\left[\varphi,\mu,\bar \mu\right]=S_L\left[\varphi,\mu,\bar \mu\right]+K\left[\mu,\bar \mu\right]\,,
\end{equation}
with
\begin{align}
	S_L\left[\varphi,\mu,\bar \mu\right]&=-\frac{c}{48\pi}\int d^2x\, \sqrt{\hat g} \left(\frac{1}{2}\hat g^{ij} \partial_i \varphi\, \partial_j \varphi+ \varphi \hat R\right)\,,\\
	K\left[\mu,\bar \mu\right]&=\frac{c}{24\pi}\int d^2z\, (1-\mu \bar \mu)^{-1}\left(\partial \mu\, \bar \partial \bar \mu-\frac{1}{2}\mu (\bar \partial \bar \mu)^2-\frac{1}{2}\bar \mu(\partial \mu)^2\right)\,.
\end{align}
It is a local functional of the metric components and thus only contributes to contact terms. It can be subtracted from the effective action $W[g_{ij}]$ in order to achieve holomorphic factorization at the expense of losing diffeomorphism invariance. The nonlocal chiral functional $W[\mu]$ is given by
\begin{equation}
\label{eq:W mu}
W[\mu]=\frac{c}{24\pi} \int d^2z\,  \frac{\bar \partial f}{\partial f}\, \partial^2 \ln \partial f\,,
\end{equation}
where the variable $f$ is implicitly related to $\mu$ through the Beltrami equation
\begin{equation}
\label{eq:Beltrami}
\mu= \frac{\bar \partial f}{\partial f}\,.
\end{equation}
The effective action $W[\mu]$ suitably generate all connected correlation functions of the chiral stress tensor component, as explicitly demonstrated in appendix~\ref{app:Yoshida}. It is also shown to satisfy the anomalous chiral diffeomorphism Ward identity,
\begin{align}
\label{eq:Ward identity main}
\left( \bar \partial- \mu \partial-2 \partial \mu \right) \frac{\delta W}{\delta \mu(z,\bz)}=\frac{c}{12\pi}\, \partial^3 \mu\,,
\end{align}
whose solution gives the well-known stress tensor expectation value in a background geometry $\mu$ related to complex plane by the (quasi)conformal mapping $f$,
\begin{equation}
\label{eq:T vev main}
\langle T \rangle_{\mu}=\pi\, \frac{\delta W[\mu]}{\delta \mu}=\frac{c}{12}\, S[f,z]\,, \qquad S[f,z]\equiv \frac{\partial^3 f}{\partial f}-\frac{3}{2}\left(\frac{\partial^2 f}{\partial f}\right)^2\,.
\end{equation}
Note that \textit{any} nonzero holomorphic function $f(z)$ is associated with a vanishing source $\mu=0$ as can be seen from \eqref{eq:Beltrami} although the stress tensor expectation manifestly depends on $f(z)$. This extra freedom relates to the existence of inequivalent but conformally related flat geometries naturally covered by the complex coordinate $w=f(z)$. To make this more explicit, we write the stress tensor in the coordinate system $w$,
\begin{equation}
\label{eq:Tww}
\langle T_{(ww)} \rangle=\left(\frac{\partial z}{\partial w}\right)^2 \langle T_{(zz)} \rangle=-\frac{c}{12}\, S[z,w]\,.
\end{equation}
In particular, the identity $f(z)=z$ yields the zero vacuum energy of the plane while $f(z)=i \ln z$ yields the constant Casimir energy of the cylinder,
\begin{equation}
\langle T \rangle_{\text{plane}}=0\,, \qquad \langle T \rangle_{\text{cyl.}}=-\frac{c}{24}\,.
\end{equation}

The effective action \eqref{eq:W mu} also arises in a slightly different context as the action for the two-dimensional quantum gravity of Polyakov in the lightcone gauge \cite{Polyakov:1987zb,Polyakov:1988qz}
\begin{equation}
	ds^2=dz d\bz+\mu(z,\bz) d\bz^2\,,
\end{equation}
where $\mu(z,\bz)$ is the only gravitational degree of freedom left after gauge fixing. This directly follows from the preceding discussion, and simply amounts to a re-interpretion of the Polyakov action \eqref{eq:Polyakov action} as describing dynamical gravity in two dimensions rather than generating the stress tensor correlations of a CFT with the metric acting as a non-dynamical background source.

It is often useful to write the effective action $W[\mu]$ in terms of the variable $F$ defined as the `inverse' of $f$,
	\begin{equation}
		\label{eq:F definition}
		F\left(f(z,\bz),\bz\right)=z\,,
	\end{equation}
	such that it takes the alternative form
	\begin{equation}
		\label{eq:W mu F}
		W[\mu]=-\frac{c}{24\pi} \int d^2w\,  \frac{\partial \bar \partial F\, \partial^2 F}{(\partial F)^2}=\frac{c}{24\pi}\int d^2w\, \frac{\bar \partial F}{\partial F}\left(\frac{\partial^3 F}{\partial F}-2 \left(\frac{\partial^2 F}{\partial F}\right)^2\right) \,,
\end{equation}
and where the integration variables are $(w,\bw)\equiv (f,\bz)$. Interestingly, this coincides with the geometric action of a particle on the vacuum coadjoint orbit of the Virasoro group \cite{Alekseev:1988ce}, where the variable $F(f)$ introduced is viewed as a coordinate on the Virasoro group while $\bw$ is interpreted as the time evolution parameter. In that context \eqref{eq:W mu F} is known as the Alekseev--Shatashvili action. 

As an aside, let me mention that the parametrization of $\mu$ in terms of a Beltrami differential is particularly well suited for the classification of inequivalent complex or conformal structures that a two-dimensional background manifold can be endowed with. More precisely, inequivalent conformal structures $\mu$ are those for which the solution $f$ to Beltrami equation \eqref{eq:Beltrami} is not continuous everywhere, in which case $f$ is called a quasiconformal mapping. The space of inequivalent conformal structures is the Teichmuller space. For more details on this and related topics, interested readers should consult the excellent reviews \cite{Alvarez:1985ez,Nelson:1986ab,Giddings:1987im,DHoker:1988pdl}.

\paragraph{From the plane to the cylinder.} The chiral generating functional $W[\mu]$ given in \eqref{eq:W mu} is defined on the plane. However, we know that correlators on the plane and cylinder are related by the conformal mapping 
\begin{equation}
z=e^{-i w}\,, \qquad w=\varphi + i \tau\,, \qquad \varphi \in [0,2\pi)\,, \quad \tau \in \mathbb{R}\,,
\end{equation} 
where $\varphi, \tau$ are coordinates covering the cylinder. It should therefore be possible to rewrite $W[\mu]$ such that it manifestly becomes the generating functional of stress tensor correlators on the cylinder. The discussion around \eqref{eq:Tww} tells us that the identity $f(z)=z$ is naturally associated with the plane while the exponential map $f(z)=i \ln z$ describes the cylinder, or respectively $F(w)=w$ and $F(w)=e^{-iw}$ in terms of the inverse variable $F$ defined in \eqref{eq:F definition}. In order to move to a description naturally suited to the cylinder, it is a good idea to apply a change of variables\footnote{Note that $\phi$ should not be confused with the Liouville field in \eqref{eq:Liouville action}.} $F \mapsto \phi$ such that the identity $\phi(w)=w$ now describes the cylinder,
\begin{equation}
\label{eq:phi definition}
F(w,\bw)= e^{-i \phi(w,\bw)}\,, \qquad \phi(\varphi+2\pi,\tau)=\phi(\varphi,\tau)+2\pi\,.
\end{equation}
Applying this change of variables to \eqref{eq:W mu F}, we obtain
\begin{equation}
\label{eq:W mu phi}
W[\mu]=-\frac{c}{24\pi} \int d^2w \left(\frac{\partial \bar \partial \phi\, \partial^2 \phi}{(\partial \phi)^2}-\bar \partial \phi\, \partial \phi \right)\,.
\end{equation}
Interestingly, this is the geometric action of a particle moving on the first exceptional coadjoint orbit of the Virasoro group \cite{Alekseev:1988ce}. In another paper \cite{Nguyen:2020jqp}, I have checked that this generating functional correctly reproduces the one- and two-point functions of the stress tensor on the cylinder, by expanding 
\begin{equation}
\phi(w,\bw)=w+\delta \phi(w,\bw)\,,
\end{equation}
and treating $\delta \phi (w,\bw)$ as an infinitesimal source for the stress tensor, but a general argument applying to all correlators for arbitrary background source geometry is still missing. For the purpose of the present work, the expression \eqref{eq:W mu phi} of the chiral generating functional will be sufficient.

Besides the plane and cylinder described above, the chiral generating functional has been given for compact Riemann surfaces of arbitrary genera in \cite{Aldrovandi:1996sa} and references therein. I will not consider these cases here.

\section{Holographic derivation in metric formulation}
\label{section:metric}
Having reviewed the generating functional of stress tensor correlations in two-dimensional CFT, we can now turn to its holographic derivation from AdS$_3$ gravity in the classical limit, i.e., in the limit of large central charge. The GKPW dictionary \eqref{GKPW classical} instructs us to compute the (suitably renormalized) onshell gravitational action with arbitrary Dirichlet conditions at the spacetime conformal boundary. Much of the material presented in this section was made available in the early days of the AdS/CFT correspondence \cite{Balasubramanian:1999re,deHaro:2000vlm}, although the story leading to the holographic Polyakov action was completed only later by Carlip \cite{Carlip:2005tz}.

The unrenormalized action that is appropriate when imposing Dirichlet conditions on the metric at a boundary surface $\partial \mathcal{M}$ is\footnote{I use  conventions where the AdS curvature is negative, which therefore differ from those adopted in~\cite{deHaro:2000vlm}.}
\begin{equation}
	\label{eq:action offshell}
	S=\frac{1}{2\kappa^2} \int_{\mathcal{M}} d^3x\, \sqrt{|G|} \left(R-2\Lambda\right)+ \frac{1}{\kappa^2} \int_{\partial \mathcal{M}} d^2x\, \sqrt{|\gamma|}\, K\,, \qquad \kappa^2=8\pi G_N\,,
\end{equation}
where the boundary surface $\partial \mathcal{M}$ is characterized by the induced metric $\gamma_{ij}$, the outward-pointing normal vector $n^\mu$ and the extrinsic curvature $K=\nabla_\mu n^\mu$. 

\paragraph{Solution space.} As is customary, we choose to write the metric in Fefferman--Graham gauge
\begin{equation}
	\label{eq:FG}
	ds^2=G_{\mu\nu}\, dx^\mu dx^\nu=\ell^2 \left(\frac{d\rho^2}{4\rho^2}+ \frac{g_{ij}(\rho,x)}{\rho}\, dx^i dx^j\right)\,,
\end{equation}
where the AdS curvature radius $\ell$ is related to the cosmological constant via $\Lambda=-1/\ell^2$. I will set $\ell=1$ in the following.
It is well-known that any solution of Einstein's equations with a negative cosmological constant locally admits an expansion in powers of $\rho$, where $\rho=0$ is the location of the spacetime conformal boundary \cite{FeffermanGraham}. In three dimensions, this expansion truncates \cite{Skenderis:1999nb}, 
\begin{equation}
g_{ij}=g^{(0)}_{ij}+\rho g^{(2)}_{ij}+\rho^2 g^{(4)}_{ij}, \qquad g^{(0)}_{ij}\equiv \tilde{g}_{ij}.
\end{equation}
In the limit where the spacetime boundary $\partial \mathcal{M}$ is the conformal boundary at infinity, the leading term $\tilde{g}_{ij}$ is fixed by the Dirichlet condition imposed on the metric field $G_{\mu\nu}$, and is interpreted as the background geometry acting as a source for the stress tensor of a dual CFT. In the following, indices $i,j$ will always be raised and lowered with $\tilde{g}_{ij}$ and its inverse. The subleading terms are partially determined by Einstein's equations,
\begin{subequations}
\begin{align}
	\label{eq:tij constraints}
	g^{(2)}_{ij}&=\frac{1}{2}\left(t_{ij}-\tilde{R}\, \tilde{g}_{ij}\right), \qquad  \tilde{\nabla}^i t_{ij}=0, \qquad \tilde{g}^{ij} t_{ij}=\tilde R,\\
	g^{(4)}_{ij}&=\frac{1}{4}g^{(2)}_{im}\, \tilde{g}^{mn}\, g^{(2)}_{nj}\,,
\end{align}
\end{subequations}
where $\tilde R$ is the Ricci curvature of $\tilde g_{ij}$. The indeterminacy in this solution is parametrized by the divergencefree tensor $t_{ij}$ whose trace is however fixed. Up to an overall constant this is just the dual CFT stress tensor in the generally covariant formulation \cite{Balasubramanian:1999re,deHaro:2000vlm}.

\paragraph{Location of the cutoff surface.}
In most of the AdS/CFT literature, the cutoff boundary surface $\partial \mathcal{M}$ is placed at a constant radial coordinate $\rho=\epsilon$, where the cutoff regulator $\epsilon$ is eventually sent to zero. However, a complete treatment that eventually yields the expected Polyakov action at the conformal boundary requires one to consider a more general `distorted' location of the cutoff boundary surface $\partial \mathcal{M}$ \cite{Carlip:2005tz}. Understanding this point requires to look at the anomalous Weyl symmetry and its holographic realization in the bulk. Since the Polyakov action ultimately comes from the Weyl anomaly, a proper treatment of Weyl transformations is crucial.

A well-known fact, premonitory of the AdS/CFT correspondence itself, is that boundary Virasoro symmetries are realized in the bulk by diffeomorphisms acting nontrivially at the cutoff boundary surface $\partial \mathcal{M}$ \cite{Brown:1986nw}. Similarly, there exists bulk diffeomorphisms, called Penrose--Brown--Henneaux (PBH) diffeomorphisms, acting as Weyl rescalings at the conformal boundary \cite{Imbimbo:1999bj,Skenderis:2000in},
\begin{equation}
\label{eq:PBH}
\tilde g'(x)=e^{2\omega (x)} \tilde g(x)\,.
\end{equation}
As reviewed in appendix~\ref{app:diffeos}, a diffeomorphism can always be traded for a change of coordinates, in this case given by \cite{Skenderis:2000in}
\begin{align}
\rho=\rho' e^{-2 \omega(x')}+\sum_{k=2}^{\infty} a_{(k)}(x') (\rho')^k\,, \qquad x^{i}=x'^{i}+\sum_{k=1}^{\infty} a^i_{(k)}(x') (\rho')^k\,.
\end{align}
All functions $a_{(k)}, a_{(k)}^i$ are completely determined from the requirement that this change of coordinates preserves the Fefferman--Graham gauge \eqref{eq:FG}, but we won't need their explicit expression. If the cutoff boundary is located at $\rho=\epsilon$, in the new coordinate system it is therefore located at
\begin{equation}
\rho'=\epsilon e^{2\omega(x')}+O(\epsilon^2)\,.
\end{equation}
Hence, one way of allowing Weyl transformations to act \textit{within} the solution space is to leave a certain freedom in the location of the cutoff surface. Dropping the primes, the location of the cutoff surface $\partial \mathcal{M}$ is chosen to satisfy
\begin{equation}
	\label{eq:boundary location}
\rho=\epsilon e^{2\omega(x)}+O(\epsilon^2) \equiv e^{2 H(x)}\,,
\end{equation}
where $\omega(x)$ is an additional degree of freedom. As it shifts under Weyl rescalings, we can rightfully expect it to be the Liouville field of the dual CFT.

Importantly, the introduction of the new degree of freedom $\omega$ renders the action principle well-defined even when the conformal factor of the boundary metric is allowed to fluctuate, i.e., when  \textit{only the conformal class of the boundary metric is kept fixed}. For this to hold $\omega$ must satisfy a boundary Liouville equation. This will be demonstrated later in this section after discussing the counterterms necessary to render the onshell action finite. As a result, PBH diffeomorphisms act within the solution space as they should.\footnote{Troessaert proposed a similar construction that differs from the present description in that the cutoff surface is kept fixed at $\rho=\epsilon$ at the expense of relaxing the FG gauge conditions  \cite{Troessaert:2013fma}.}  

\begin{figure}
	\centering
	\includegraphics[scale=0.7]{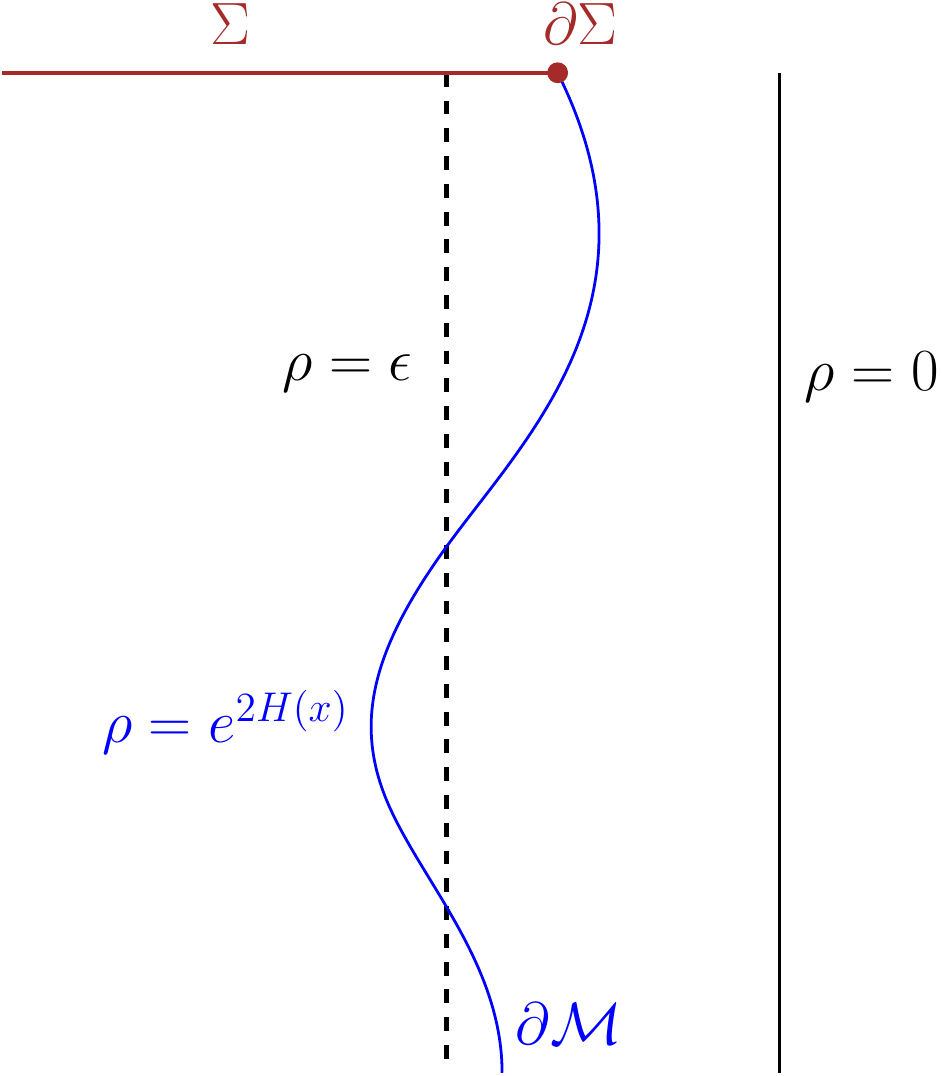}
	\caption{Schematic representation of the gravitational setup. The manifold $\mathcal{M}$ under consideration is bounded by a cutoff boundary surface $\partial \mathcal{M}$ close to spatial infinity ($\rho=0$), and by potential Cauchy surfaces $\Sigma$. The cutoff surface at $\rho=e^{2H(x)}$ is the image of the surface at $\rho=\epsilon$ under a Penrose--Brown--Henneaux diffeomorphism.}
	\label{figure}
\end{figure}

\paragraph{Onshell action.}
We can now evaluate the onshell action. Using Einstein's equations and explicitly performing the $\rho$ integration, the bulk term reduces to
\begin{subequations}
\label{eq:log divergence}
\begin{align}
	\frac{1}{2\kappa^2} \int_{\mathcal{M}} d^3x\, \sqrt{|G|} \left(R-2\Lambda\right)&=-\frac{1}{\kappa^2} \int_{\rho \geq e^{2F(x)}} d^3x\, \sqrt{|g|}\, \rho^{-2}\\
	&=-\frac{1}{\kappa^2} \int_{\rho \geq e^{2F(x)}} d^3x\, \sqrt{|\tilde g|} \left[\rho^{-2}-\frac{\tilde R}{4\rho}+ O(\rho^0)\right]\\
	&=\frac{1}{\kappa^2} \int d^2x\, \sqrt{|\tilde g|} \left[-\rho^{-1}-\frac{1}{4}\log \rho\, \tilde R+ O(\rho)\right]_{\rho=e^{2H(x)}}\,.
\end{align}
\end{subequations}
The normal vector to the cutoff surface $\partial \mathcal{M}$ characterized by the equation \eqref{eq:boundary location}, is explicitly given by
\begin{subequations}
\begin{align}
	n^\rho&=-2\rho \left(1+G^{ij}\partial_i H \partial_j H\right)^{-1/2}=-2\rho+O(\rho^2)\,,\\
	n^i&=G^{ij}\partial_jH \left(1+G^{ij}\partial_i H \partial_j H\right)^{-1/2}=\rho\, \tilde{g}^{ij} \partial_j H+O(\rho^2)\,,
\end{align}
\end{subequations}
such that the extrinsic curvature can be expressed as
\begin{equation}
\label{eq:K}
K=\nabla_\mu n^\mu=\frac{1}{\sqrt{|G|}}\, \partial_\mu \left(\sqrt{|G|}\,  n^\mu\right)=2+\rho \left( \tilde \square H+\frac{1}{2}\tilde R\right)+O(\rho^2)\,.
\end{equation}
On the other hand, the induced metric $\gamma_{ij}$ at $\partial \mathcal{M}$ reads
\begin{subequations}
\begin{align}
	\gamma_{ij}&=G_{ij}+\partial_i H \partial_j H=\rho^{-1} \tilde{g}_{ij}+ g^{(2)}_{ij}+\partial_i H \partial_j H+O(\rho)\,,\\
	\sqrt{|\gamma|}&=\sqrt{|\tilde g|}\left(\rho^{-1}+\frac{1}{2} (\partial H)^2-\frac{1}{4} \tilde R+O(\rho)\right)\,.
\end{align}
\end{subequations}
Plugging this back into the action \eqref{eq:action offshell}, we find
\begin{equation}
S_{\reg}=\frac{1}{2\kappa^2} \int d^2x\, \sqrt{|\tilde g|} \left[2 \rho^{-1}-\frac{1}{2} \ln \rho\, \tilde R+2 (\partial H)^2+2 \tilde \square H+O(\rho)\right]_{\rho=e^{2H}}\,.
\end{equation}
The power-law divergence associated with the first term is eliminated by the addition of a boundary counterterm to the action \cite{Balasubramanian:1999re},
\begin{equation}
\label{eq:counterterm}
S_{\ct}=-\frac{1}{\kappa^2} \int_{\partial \mathcal{M}} d^2x\, \sqrt{|\gamma|}\,.
\end{equation}
This counterterm is a local functional of the boundary intrinsic geometry, and therefore does not alter the variational principle with Dirichlet conditions imposed on $\gamma_{ij}$. This is interpreted as a UV regularization in the dual theory, which preserves full boundary diffeomorphism invariance. Expanding the function $H(x)$ in powers of the cutoff regulator $\epsilon$ as in \eqref{eq:boundary location}, we finally obtain
\begin{subequations}
\begin{align}
	S_{\reg}+S_{\ct}&=\frac{1}{2\kappa^2} \int d^2x\, \sqrt{|\tilde{g}|} \left[-H \tilde R + (\partial H)^2+\frac{1}{2} \tilde R +2\tilde \square H+O(\rho)\right]\\
	&=\frac{1}{2\kappa^2} \int d^2x\, \sqrt{|\tilde g|} \left[-\frac{1}{2} \ln \epsilon\, \tilde R-\omega \tilde R + (\partial \omega)^2+\frac{1}{2} \tilde R +2 \tilde \square \omega+O(\epsilon)\right]\,.
\end{align}
\end{subequations}
It may look like there is still a logarithmic UV divergence in the limit $\epsilon \to 0$, but the integral of the Ricci curvature is topological in two dimensions and we simply discard it, since we haven't been careful about boundary and corner terms at $\Sigma$ and $\partial \Sigma$ anyways (see figure~\ref{figure}). In higher dimensions, one would need to introduce a logarithmically divergent counterterm that depends explicitly on the cutoff regulator $\epsilon$. Counterterms of this sort explicitly break Weyl invariance, a fact closely related to the appearance of a Weyl or conformal anomaly. Similarly dropping the total derivative term, we end up with the \textit{renormalized} action
\begin{equation}
\label{eq:S ren}
S_{\ren}=\lim\limits_{\epsilon \to 0}\left(S_{\reg}+S_{\ct}\right)=\frac{1}{2\kappa^2} \int d^2x\, \sqrt{|\tilde g|} \left[(\partial \omega)^2-\omega \tilde R\right]\,.
\end{equation} 
We recognize the Liouville action \eqref{eq:Liouville action} that we were after, with $\omega$ playing the role of the Liouville field. The action is not yet fully onshell since $\omega$ is a dynamical mode not constrained by any sort of boundary condition. Solving its equation of motion $2\tilde \square \omega =-\tilde R$, and plugging it back into the action yields
\begin{equation}
\label{eq:W metric}
W[\tilde g_{ij}]\equiv S_{\ren}^{\text{onshell}}=\frac{c}{96\pi} \int d^2x\, \sqrt{|\tilde g|}\, \tilde R\,  \tilde{\square}^{-1} \tilde R\,,
\end{equation}
where we have replaced Newton's constant by the central charge using \eqref{c}. This is exactly the expected result, which explicitly demonstrates the validity of the GKPW dictionary \eqref{GKPW classical}. Note that the Liouville equation
\begin{equation}
\label{eq:Liouville omega}
\tilde \square \omega =-\frac{1}{2}\tilde R\,, 
\end{equation}
in fact directly follows from the requirement of a well-posed action principle when only the \textit{conformal class} of metrics $[\gamma_{ij}]$ is kept fixed at the boundary. Indeed, allowing boundary variations of the form $\delta \gamma_{ij}=\delta \sigma\, \gamma_{ij}$ and imposing Einstein's equations, we find 
\begin{equation}
\delta \left(S+S_{\ct}\right)=\frac{1}{2\kappa^2} \int_{\partial \mathcal{M}} d^2x\, \sqrt{|\gamma|} \left(2-K\right)\delta \sigma\,.
\end{equation}  
Looking at the explicit expression for $K$ given in \eqref{eq:K}, we see that the Liouville equation \eqref{eq:Liouville omega} is precisely what we need to enforce in order to have a well-posed action principle with arbitrary $\delta \sigma$, i.e., when only the conformal class $[\gamma_{ij}]$ is kept fixed. This is a highly desirable feature since the conformal classes $[\tilde g_{ij}]$ are the only available structures at infinity. It also fits with the holographic picture where the objects of interest are conformal field theories defined at conformal infinity. The problem of finding such an action principle is relatively old and has been discussed in several papers without complete success \cite{Papadimitriou:2005ii,Troessaert:2013fma,Alessio:2020ioh}. We see that it is realized for free in the present setup, which incidentally yields the correct stress tensor generating functional.

It is also worth contrasting the above derivation due to Carlip to some earlier classic work in which the conformal mode $\omega$ was not taken into account (see e.g.~\cite{deHaro:2000vlm}). Indeed, setting $\omega(x)=0$ such that the boundary cutoff surface lies at a fixed radial coordinate $\rho=\epsilon$, one finds the renormalized onshell action
\begin{equation}
\label{naive Sren}
S_{\ren}^{\text{naive}}=\frac{1}{4\kappa^2} \int d^2x\, \sqrt{|\tilde{g}|}\, \tilde{R} \,.
\end{equation}
This cannot be correct however, since this yields a vanishing stress tensor vacuum expectation value (vev)
\begin{equation}
\label{eq:T naive}
\langle T_{ij} \rangle^{\text{naive}} \sim \frac{\delta S_{\ren}^{\text{naive}}}{\delta \tilde g^{ij}} \sim \tilde{R}_{ij}-\frac{1}{2} \tilde{R}\, \tilde{g}_{ij}=0\,,
\end{equation}
for any two-dimensional metric $\tilde{g}_{ij}$. As already emphasized, the integral \eqref{naive Sren} is topological in two dimensions. In particular, \eqref{eq:T naive} does not reproduce the trace anomaly. Note that the stress tensor vev is successfully derived in \cite{deHaro:2000vlm}, however not by straightforward application of the GKPW dictionary \eqref{GKPW classical}. Indeed, the authors compute it via
\begin{equation}
\langle T_{ij} \rangle \sim \int d^2x\, \frac{\delta (S_{\reg}+S_{\ct})}{\delta G^{\mu\nu}} \frac{\delta G^{\mu\nu}}{\delta \tilde g^{ij}} \Big|_{\text{onshell}} \sim c\, t_{ij}\,,
\end{equation}
where arbitrary field variations are applied \textit{before} onshell evaluation. Although the difference is subtle, this is not equivalent to evaluating the action onshell and only then take its variation with respect to the boundary metric.

\section{Holographic derivation in Chern--Simons formulation}
\label{section:Chern-Simons}
In this section I present a holographic derivation of the Polyakov action in the Chern--Simons formulation of three-dimensional gravity with negative cosmological constant. This derivation will closely follow the one given in section~\ref{section:metric} in the metric formalism. In particular, manifest covariance with respect to the boundary geometry will be kept at all times. Importantly, this derivation is not equivalent to the classic Hamiltonian reduction to a boundary flat Liouville theory \cite{Coussaert:1995zp}, nor to the more recent derivation of two chiral copies of the Alekseev--Shatashvili action \cite{Cotler:2018zff}. I will further comment on these various approaches in sections~\ref{section:Hamiltonian reduction} and \ref{section:Cotler}.

\paragraph{Chern--Simons gravity.} 
The Chern--Simons formulation of three-dimensional gravity was introduced in \cite{Achucarro:1987vz,Witten:1988hc}. To properly describe it, we need a first order formulation of gravity in tetrad and spin connection variables. The tetrad $\ea_\mu$ with indices $\sA=0,1,2$ is a set of local orthonormal frame fields satisfying 
\begin{equation}
\label{eq:G from ea}
G_{\mu\nu}=\eta_{\sA\sB} \ea_\mu \eb_\nu\,,
\end{equation}
where $\eta_{\sA \sB}$ is the flat Lorentzian metric of the tangent space,
\begin{equation}
\eta_{\sA\sB}=
\begin{pmatrix}
0 & 1 & 0\\
1 & 0 & 0\\
0 & 0 & 1
\end{pmatrix}\,.
\end{equation}
Indices are raised and lowered with this metric. The antisymmetric spin connection one-form $\omega^{\sA \sB}=-\omega^{\sB \sA}$ is a connection that is compatible with the metric $G_{\mu\nu}$,  
\begin{equation}
\nabla_\mu \ea_\nu=-\omega\indices{_\mu^\sA_\sB}\, \eb_\nu\,,
\end{equation}
with curvature two-form given by
\begin{equation}
R^{\sA \sB}=d\omega^{\sA \sB}+\omega\indices{^\sA_\sC} \wedge \omega^{\sC \sB}\,.
\end{equation}
In terms of these variables, the bulk Einstein--Hilbert action can be written
\begin{equation}
\label{eq:EH action}
S_{\text{EH}}=\frac{1}{2\kappa^2} \int_{\mathcal{M}} \epsilon_{\sA \sB \sC} \left( \ea \wedge R^{\sB \sC}-\frac{\Lambda}{3}\, \ea \wedge \eb \wedge \ec \right)\,,
\end{equation}
where $\epsilon_{\sA \sB \sC}$ is the totally antisymmetric Levi-Civita symbol with $\epsilon_{012}=1$. The resulting equations of motion are Einstein's equations together with the vanishing of the torsion
\begin{equation}
T^{\sA}\equiv d\ea+\omega\indices{^{\sA}_{\sB}} \wedge \eb=0\,,
\end{equation}
which can be used to algebraically solve $\omega^{\sA \sB}$ in terms of the tetrad $\ea$. A crucial feature in three dimensions is that we can dualize the spin connection,
\begin{equation}
\omega_{\sA} \equiv \frac{1}{2}\, \epsilon_{\sA \sB \sC}\, \omega^{\sB \sC} \qquad \longleftrightarrow \qquad \omega^{\sA \sB}=-\epsilon^{\sA \sB \sC}\, \omega_{\sC}\,.
\end{equation}
We can then introduce two chiral Chern--Simons connection valued in the $\mathfrak{sl}(2,\mathbb{R})=\mathfrak{so}(2,1)$ algebra,
\begin{equation}
\label{eq:A definitions}
A=A^{\sA} \Jt_{\sA}=\left(\oa+\frac{\ea}{\ell} \right) \Jt_{\sA}\,, \qquad \bA=\bA^{\sA} \bJt_{\sA}=\left(\oa-\frac{\ea}{\ell} \right) \bJt_{\sA}\,, 
\end{equation}
where the algebra generators are taken to satisfy
\begin{equation}
\left[\Jt_{\sA},\Jt_{\sB}\right]=\epsilon_{\sA \sB \sC} \Jt^{\sA}\,, \qquad \Tr\left[\Jt_{\sA}\Jt_{\sB}\right]=\frac{1}{2} \eta_{\sA \sB}\,,
\end{equation}
and can be represented by the matrices
\begin{align}
\label{eq:J matrices}
\Jt_0=\frac{1}{\sqrt{2}} 
\begin{pmatrix}
0 & 1\\
0 & 0
\end{pmatrix}\,, \qquad \Jt_1=\frac{1}{\sqrt{2}} 
\begin{pmatrix}
0 & 0\\
1 & 0
\end{pmatrix}\,, \qquad \Jt_2=\frac{1}{2} 
\begin{pmatrix}
1 & 0\\
0 & -1
\end{pmatrix}\,.
\end{align}
The Einstein--Hilbert action \eqref{eq:EH action} essentially coincides with the difference of two chiral $\mathfrak{sl}(2,\mathbb{R})$ Chern--Simons actions, 
\begin{equation}
\label{eq:CS to EH}
S_{\CS}[A]-S_{\CS}[\bA]=S_{\text{EH}}+\frac{1}{2\kappa^2} \int_{\partial \mathcal{M}} \ea \wedge \omega_{\sA}\,,
\end{equation}
with 
\begin{equation}
S_{\CS}[A]=\frac{k}{4\pi} \int \Tr[A\wedge dA+\frac{2}{3}\, A\wedge A \wedge A ]\,, \qquad k=\frac{\ell}{4G_N}\,.
\end{equation}
The resulting equations of motion are the vanishing of the field-strengths,
\begin{equation}
\label{eq:eom CS}
F[A]=F[\bA]=0\,, \qquad F[A]=dA+A \wedge A\,.
\end{equation}
They are equivalent to Einstein's equations and the torsionfree constraint written in first order variables.

\paragraph{Boundary geometry in tetrad variables.}
As described above, there is a one-to-one correspondence between the CS connections $A, \bA$ and the first order variables $\ea, \oa$. It is therefore natural to also describe the boundary geometry in terms of a two-dimensional tetrad $\tea$ with index $\sa=0,1$ and associated (dualized) spin connection one-form $\tomega$ satisfying
\begin{equation}
\tilde g_{ij}=\te^0_i\, \te^1_j+\te^0_j\, \te^1_i\,, 
\end{equation}
and
\begin{equation}
d \te^0+ \tomega \wedge \te^0=0\,, \qquad d \te^1-\tomega \wedge \te^1=0\,.
\end{equation}
The last set of equation is the requirement that the spin connection $\tomega$ be torsionfree. In terms of the spin connection, the scalar curvature is given by
\begin{equation}
\tilde R=-2 \tilde \varepsilon^{ij} \partial_i \tomega_j\,.
\end{equation}

\subsection{Covariant approach}
\label{section:Banados}
I will now describe the emergence of a Polyakov action at the conformal boundary in a way that closely parallels the derivation in the metric formalism described in section~\ref{section:metric}. Since the Polyakov action is manifestly covariant with respect to the boundary metric $\tilde g_{ij}$, it is best to keep this covariance manifest throughout our computations. In order to achieve this, I will adopt the covariant description of Ba\~nados, Chandia and Ritz \cite{Banados:2002ey}. We choose coordinates $(r,x^i)$ where $r$ is a radial coordinate approaching $r \to \infty$ at the conformal boundary and $x^i$ are arbitrary coordinates on constant-$r$ surfaces. For our purposes, it will be sufficient to consider an ansatz of the form
\begin{subequations}
\label{eq:radial ansatz}
\begin{align}
	A&=h^{-1} a h+h^{-1} d h\,, \qquad h=e^{r \Jt_2}\,,\\
	\bA&=\bh^{-1} \ba \bh+\bh^{-1} d \bh\,, \qquad \bh=e^{-r \Jt_2}\,.
\end{align}
\end{subequations}
where $a, \ba$ are purely transverse and $r$-independent connections. This ansatz automatically solves the `constraints' $F_{ri}=0$. Using $e^{-r \Jt_2} \Jt_0 e^{r \Jt_2}= e^{- r} \Jt_0$ and $e^{-r \Jt_2} \Jt_1 e^{r \Jt_2}= e^{r} \Jt_1$, \eqref{eq:radial ansatz} reduces to
\begin{subequations}
\label{eq:A expansion}
\begin{align}
	A&=e^{-r} a^0 \Jt_0 +e^{r} a^1 \Jt_1 +\left(a^2+dr\right)  \Jt_2\,,\\
	\bA&=e^{r} \ba^0 \bJt_0 +e^{-r} \ba^1 \bJt_1 +\left(\ba^2-dr\right)  \bJt_2\,.
\end{align}
\end{subequations}
The associated bulk metric $G_{\mu\nu}$, obtained after extracting $\ea$ from \eqref{eq:A definitions} and making use of \eqref{eq:G from ea}, contains a cross terms $dr dx^i$ unless $a^2=\ba^2$, which we choose to impose as an additional restriction. In this case the tetrad and spin connection are found to be
\begin{align}
	\label{eq:e expansion}
e^0=\frac{1}{2}\left(e^{-r} a^0 -e^{r} \ba^0\right)\,, \qquad e^1=\frac{1}{2}\left(e^{r} a^1 -e^{-r} \ba^1\right)\,, \qquad
e^2=dr\,,
\end{align}
and
\begin{equation}
	\label{eq:omega expansion}
\omega^0=\frac{1}{2}\left(e^{-r} a^0 +e^{r} \ba^0\right)\,, \qquad \omega^1=\frac{1}{2}\left(e^{r} a^1 +e^{-r} \ba^1\right)\,, \qquad
\omega^2=\frac{1}{2}\left(a^2+\ba^2\right)\,.
\end{equation}
Introducing the radial coordinate $\rho=e^{-2r}$, the associated bulk metric is
\begin{equation}
\label{eq:CS metric}
ds^2=2 e^0 e^1+e^2e^2=\frac{d\rho^2}{4\rho^2} +\frac{1}{2}\left(-\rho^{-1}\, \ba^0 a^1 +a^0 a^1+\ba^0 \ba^1-\rho\, a^0 \ba^1\right)\,.
\end{equation}
In order to make the correspondence with the Fefferman--Graham expansion \eqref{eq:FG} more explicit, we identify the boundary tetrad and spin connection
\begin{equation}
\label{eq:identification 1}
\te^0= -\frac{\ba^0}{2}\,, \qquad \te^1=\frac{a^1}{2}\,, \qquad \tomega = a^2=\ba^2\,,
\end{equation}
together with components of the subleading term in the metric, 
\begin{equation}
\label{eq:identification 2}
g^{(2)\sa}_i\equiv g\indices{^{(2)}_i^j} \tea_j\,, \qquad g^{(2)0}=a^0\,, \qquad g^{(2)1}=-\ba^1\,.
\end{equation}
The bulk metric \eqref{eq:CS metric} then takes the form 
\begin{equation}
ds^2=\frac{d\rho^2}{4\rho^2} + \left(\rho^{-1} \tilde g_{ij} + g^{(2)}_{ij}+\frac{\rho}{4}g^{(2)}_{ik} g\indices{^{(2)k}_j}\right)dx^i dx^j\,.
\end{equation}
This already almost exactly describes the solution space of pure gravity in Fefferman--Graham gauge, except that we still have to derive the constraints on $g^{(2)}$. These will follow from the equations of motion.

\paragraph{Solution space.} Starting from the ansatz \eqref{eq:radial ansatz}, the equations of motion \eqref{eq:eom CS} are
\begin{subequations}
\begin{align}
	d a^1-a^2\wedge a^1&=0\,, \qquad d \ba^0 +a^2 \wedge \ba^0=0\,,\\
	d a^0 +a^2 \wedge a^0&=0\,, \qquad d \ba^1 -a^2 \wedge \ba^1=0 \,,\\
	\label{eq:da2}
	d a^2+a^0 \wedge a^1&=0\,, \qquad d a^2+\ba^0 \wedge \ba^1=0\,,
\end{align}
\end{subequations}
which are just $da+a \wedge a=0=d\ba+ \ba \wedge \ba$ expanded in $\mathfrak{sl}(2,\mathbb{R})$ basis. With the identifications made in \eqref{eq:identification 1}-\eqref{eq:identification 2}, the first set of equations can be rewritten
\begin{align}
d \te^0+ \tomega \wedge \te^0=0\,, \qquad d \te^1-\tomega \wedge \te^1=0\,.
\end{align}
This means that $\tomega$ is indeed the torsionfree connection compatible with the metric $\tilde g_{ij}$. The remaining equations can be combined into
\begin{subequations}
\begin{gather}
	d\tomega +\epsilon_{\sa \sb}\, \te^{\sa} \wedge g^{(2)\sb}=0\,, \\
	dg^{(2)\sa}-\epsilon\indices{^{\sa}_{\sb}}\, \tomega \wedge g^{(2)\sb}=0\,.
\end{gather}
\end{subequations}
Writing
\begin{equation}
	g^{(2)\sa}=\frac{1}{2}\left(t^{\sa}-\tilde R \tea \right)\,, 
\end{equation}
we recover the trace and divergence constraints \eqref{eq:tij constraints} on the holographic stress tensor, expressed in frame components,
\begin{equation}
d t^{\sa}-\epsilon^{\sa \sb} \tomega \wedge t_{\sb}-d\tilde R \wedge \tea=0\,, \qquad \te_{\sa}^i t^{\sa}_i=\tilde R\,.
\end{equation}
Thus far we find perfect agreement with the metric formulation.

\paragraph{Action principle.} We now turn to the definition of a suitable action principle for the Chern--Simons theory that will yield the desired solution space described above. We want to treat the induced boundary geometry as a fixed but otherwise arbitrary background source for the dual CFT. This is achieved by imposing the following boundary conditions on the CS connections at $\partial \mathcal{M}$,
\begin{equation}
\label{eq:CS bc}
\delta(A^0 - \bA^0)=\delta(A^1 - \bA^1)=\delta(A^2+\bA^2)=0\,, \qquad (a^2=\ba^2)\,.
\end{equation} 
Taken together, these boundary conditions fix the induced boundary geometry $\lbrace \tea\,, \tomega\rbrace $ at the cutoff surface $\partial \mathcal{M}$. 
The appropriate action to consider is then \cite{Banados:2002ey}
\begin{equation}
\label{eq:CS total action}
S[A,\bA]=S_{\CS}[A]-S_{\CS}[\bA]+\frac{k}{8\pi} \int_{\partial \mathcal{M}} A^0 \wedge \bA^1 + A^1 \wedge \bA^0-A^2 \wedge \bA^2\,.
\end{equation}
Note that the boundary term $A^0 \wedge \bA^1$ actually vanishes in the limit where the cutoff boundary $\partial \mathcal{M}$ is sent to infinity, as can be seen from \eqref{eq:A expansion}. Onshell variation of the action yields
\begin{align}
&\delta S[A,\bA]\\
\nonumber
&=-\frac{k}{8\pi} \int_{\partial \mathcal{M}} (A^0+\bA^0) \wedge \delta (A^1-\bA^1) + (A^1+\bA^1) \wedge \delta (A^0-\bA^0)+(A^2-\bA^2) \wedge \delta (A^2+\bA^2)\,,
\end{align}
and we conclude that the action is stationary for bulk solutions satisfying the boundary conditions \eqref{eq:CS bc}.
	
\paragraph{Onshell action.} As in the metric formalism, we want to evaluate the onshell action as we later identify it with the generating functional of the dual CFT after appropriate renormalization. Direct evaluation of the bulk Lagrangian gives
\begin{equation}
\Tr[A \wedge dA +\frac{2}{3} A \wedge A \wedge A]=\Tr\left[da \wedge dr \Jt_2\right]=d\left(\Tr\left[r da \Jt_2\right]\right)=\frac{1}{2}d\left(r d\tomega\right)\,,
\end{equation}	
such that
\begin{subequations}
\label{eq:CS bulk onshell}
\begin{align}
	S_{\CS}[A]&=\frac{k}{8\pi} \int_{\partial \mathcal{M}} r d\tomega=-\frac{k}{8\pi} \int_{\partial \mathcal{M}} d^2x\, \sqrt{-\tilde g}\, r \varepsilon^{ij} \partial_i \tomega_j=\frac{k}{16\pi} \int_{\partial \mathcal{M}} d^2x\, \sqrt{|\tilde g|}\, r \tilde R\,,\\
	S_{\CS}[\bA]&=-S_{\CS}\left[A\right]\,.
\end{align}
\end{subequations}
At this point it is useful to check the consistency between the computations done in metric and Chern--Simons formulations. According to \eqref{eq:CS to EH}, the difference between $S_{\CS}[A]-S_{\CS}[\bA]$ and $S_{\text{EH}}$ should be given by
\begin{equation}
\label{eq:extra term}
\frac{1}{2\kappa^2} \int_{\partial \mathcal{M}} \ea \wedge \omega_{\sA}=\frac{1}{2\kappa^2} \int_{\partial \mathcal{M}} 2e^{2r}\, \te^0 \wedge \te^1 - r d\tomega=\frac{1}{2\kappa^2} \int_{\partial \mathcal{M}} d^2x\, \sqrt{|\tilde g|}\, \left(2e^{2r} -\frac{r}{2} \tilde R\right)\,.
\end{equation} 
Recalling that $\rho=e^{-2r}$ and putting together \eqref{eq:log divergence}, \eqref{eq:CS bulk onshell} and \eqref{eq:extra term}, we find that equation \eqref{eq:CS to EH} is indeed satisfied. On the other hand, evaluation the boundary term in \eqref{eq:CS total action} simply yields 
\begin{subequations}
\begin{align}
A^0 \wedge \bA^1 + A^1 \wedge \bA^0-A^2 \wedge \bA^2&=e^{2r} a^1 \wedge \ba^0-dr \wedge \tomega +O(e^{-2r})\\
&=4 e^{2r} \te^0 \wedge \te^1+r d\tomega -d(r \tomega) +O(e^{-2r})\,.
\end{align}
\end{subequations}
Therefore, up to corner terms at $\partial \Sigma$ the total regulated onshell action is
\begin{equation}
S_{\reg}=\frac{k}{4\pi} \int_{\partial \mathcal{M}} d^2x\, \sqrt{|\tilde g|} \left[2 e^{2r}\, +r \tilde R+O(e^{-2r})\right]\,.
\end{equation}
Like in the metric formalism, the leading divergence is cured by adding the intrinsic action counterterm \eqref{eq:counterterm}. Evaluating the integrand at the boundary location $r=-H(x)=-\frac{1}{2} \ln \epsilon-\omega(x)+O(\epsilon)$ and taking the limit $\epsilon \to 0$, we end up with the renormalized onshell action 
\begin{equation}
\label{eq:CS Liouville}
S_{\ren}=\lim\limits_{\epsilon \to 0}\left(S_{\reg}+S_{\ct}\right)=-\frac{k}{4\pi} \int d^2x\, \sqrt{|\tilde g|}\, \left[(\partial \omega)^2+\omega \tilde R\right]\,,
\end{equation} 
where topological terms have again been dropped. This slightly differs from the result \eqref{eq:S ren} in metric formalism, but we are not done yet. As before, the generating functional $W[\tilde g_{ij}]$ given in \eqref{eq:W metric} is obtained after elimination of the Liouville field $\omega$, yielding
\begin{equation}
\label{eq:W CS}
W[\tilde g_{ij}]\equiv S_{\ren}^{\text{onshell}}=\frac{c}{96\pi} \int d^2x\, \sqrt{|\tilde g|}\, \tilde R\,  \tilde{\square}^{-1} \tilde R\,.
\end{equation}
The Polyakov action is once again recovered. Note that the agreement with the result of section~\ref{section:metric} is not completely trivial since the boundary terms used in metric (second order) and Chern--Simons (first order) formulations differ.

\subsection{Standard Hamiltonian reduction}
\label{section:Hamiltonian reduction}
It is worth contrasting the above derivation of the holographic generating functional with the classic Hamiltonian reduction of three-dimensional gravity to a flat Liouville theory with nonzero potential \cite{Coussaert:1995zp}. See also the detailed review \cite{Donnay:2016iyk}.

\paragraph{Restricted phase space.} The setup of the Hamiltonian reduction crucially differs from that of the preceding subsection in that the boundary geometry is taken to be flat. Sources for the dual stress tensor are turned off. Also, the boundary $\partial \mathcal{M}$ is taken to lie a constant $r$-surface, i.e, $\omega(x)=0$. As a preliminary step, a particular coordinate system $(r,t,\varphi)$ is chosen where $t$ is timelike and $\varphi$ is spacelike and periodic. Coordinates $z=t+\varphi$ and $\bz=t-\varphi$ will also be used. Still adopting the general ansatz \eqref{eq:A expansion}, the purely transverse connection is restricted to the form 
\begin{subequations}
\label{eq:a restricted}
\begin{align}
a&=\sqrt{2}\left(L(z,\bz)\Jt_0+\Jt_1\right) dz\,,\\
\ba&=\sqrt{2}\left(\Jt_0+\bL(z,\bz)\Jt_1\right) d\bz\,,
\end{align}
\end{subequations}
such that the bulk metric reads
\begin{equation}
\label{eq:flat solutions}
ds^2=\frac{d\rho^2}{4\rho^2}- \rho^{-1} \left(dz- \rho\, \bL(z,\bz) d\bz \right)\left(d\bz - \rho\, L(z,\bz) dz \right)\,.
\end{equation}
The boundary geometry of this restricted ansatz is clearly flat. Note also that is not a classical solution unless $\bar \partial L=0=\partial \bL$.

\paragraph{WZNW model.}  The chiral Chern--Simons action takes the explicit form
\begin{align}
\label{eq:S hamiltonian}
S_{\CS}[A]=\frac{k}{4\pi}\int_{\mathcal{M}} d^3x\,  \Tr[\partial_r(A_\varphi A_t)+A_r \dot{A}_\varphi-A_\varphi \dot{A}_r-2A_t F_{r\varphi}]\,,
\end{align}
where dots refer to $\partial_t$ derivatives, while primes will refer to $\partial_\varphi$ derivatives.
Onshell variation of this action yields 
\begin{equation}
\delta S_{\CS}[A]=-\frac{k}{4\pi}\int_{\partial \mathcal{M}} dt d\varphi\,  \Tr[A_t\,  \delta A_\varphi - A_\varphi\, \delta A_t]\,.
\end{equation}
The boundary conditions adopted by Coussaert, Henneaux and van Driel is $A_t=A_\varphi$ at $\partial \mathcal{M}$, which makes $S_{\CS}[A]$ stationary without any additional boundary term. On the connection $\bA$ one imposes $\bA_t=-\bA_\varphi$ at $\partial \mathcal{M}$. The restricted phase-space \eqref{eq:a restricted} does satisfy these conditions. Importantly, they are completely distinct from the boundary conditions \eqref{eq:CS bc} used in the preceding subsection to allow fixed but otherwise arbitrary boundary geometries. 

The component $A_t$ manifestly plays the role of a Lagrange multiplier for the constraint $F_{r\varphi}=0$, solved by
\begin{equation}
\label{eq:solution constraint}
A_r=G^{-1} \partial_r G\,, \qquad A_\varphi=G^{-1} \partial_\varphi G\,, \qquad \forall\, G \in \text{SL}(2,\mathbb{R})\,.
\end{equation}
Here I have assumed trivial spacetime topology such that $A$ and $\bA$ do not possess holonomies. On the constraint surface $F_{r\varphi}=0$, the quantity
\begin{equation}
\Tr[A \wedge dA+\frac{2}{3} A\wedge A \wedge A]-d^3x\,  \Tr\left[\partial_r(A_\varphi A_t)\right]
\end{equation}
is actually completely independent of the value taken by $A_t$. Without any loss of generality, it can therefore be evaluated using 
\begin{equation}
A=G^{-1}dG\,, \qquad (dA+A \wedge A=0)\,.
\end{equation}
This allows to directly express the action in terms of the group element $G$ as
\begin{align}
S_{\CS}[A]=-\frac{k}{12\pi}\int_{\mathcal{M}} \Tr\left[(G^{-1}dG)^3\right]+ \frac{k}{4\pi} \int_{\partial \mathcal{M}} dt d\varphi\, \Tr\left[\left(A_t -G^{-1}\partial_t G\right) G^{-1}\partial_\varphi G\right]\,.
\end{align}
Using the boundary condition $A_t=A_\varphi=G^{-1} \partial_\varphi G$, this reduces to a chiral Wess--Zumino--Novikov--Witten (WZNW) model
\begin{equation}
\label{eq:WZNW}
S_{\CS}\left[A\right]=-\frac{k}{12\pi} \int \Tr\left[(G^{-1}dG)^3\right]-\frac{k}{2\pi}\int dt d\varphi\, \Tr\left[G^{-1} \bar \partial G\, G^{-1} \partial_\varphi G\right]\,.
\end{equation}	

\paragraph{Flat Liouville theory.} 
We now come to the well-known reduction of the chiral WZNW actions \eqref{eq:WZNW} to that of a flat Liouville theory with nonzero potential. Since the boundary geometry is flat, this Liouville theory cannot possibly be related to the one described in previous sections. In fact, we will see that it is still off-shell and further vanishes when the bulk equations of motion are imposed.

In the original work of Coussaert, Henneaux and Van Driel, the Liouville theory was obtained by first combining the chiral WZNW models into a single non-chiral WZNW model through field redefinitions. These field redefinitions make it very difficult to interpret the result in terms of gravitational variables and corresponding boundary conditions. Fortunately, another derivation of the same result has been given in \cite{Henneaux:1999ib,Barnich:2013yka} which sidesteps the non-chiral WZNW model. Here, I simply reproduce the computations presented in \cite{Barnich:2013yka}. As a first step, we write
\begin{equation}
G=g(t,\varphi) \cdot h(r)\,,
\end{equation}
with $h$ given in \eqref{eq:radial ansatz}, such that 
\begin{equation}
a_t=a_\varphi=g^{-1} \partial_\varphi g\,.
\end{equation}
We then write a Gauss parametrization of the group element $g$, 
\begin{equation}
\label{eq:Gauss main}
g=e^{\sqrt{2} \sigma \Jt_1} e^{-\phi \Jt_2} e^{\sqrt{2} \tau \Jt_0}\,,
\end{equation}
where $\sigma, \phi, \tau$ are functions of the boundary coordinates $z,\bz$. With this decomposition, the transverse connection $a$ reads
\begin{align}
\label{eq:at aphi}
a_t=a_\varphi=\sqrt{2}\left(\tau'- \tau \phi'-e^{-\phi} \tau^2  \sigma'\right) \Jt_0 + \sqrt{2} e^{-\phi} \sigma' \Jt_1-\left( \phi'+2 e^{-\phi}\tau \sigma'\right)\Jt_2\,.
\end{align}
In turn, the ansatz \eqref{eq:a restricted} imposes the boundary conditions
\begin{align}
\label{eq:bc Liouville}
\sigma'=e^\phi\,, \qquad \phi'=-2\tau\,,
\end{align}
while the free function $L$ is identified with
\begin{equation}
L(z,\bz)=\tau' +\tau^2=\frac{1}{4}(\phi')^2-\frac{1}{2}  \phi''=-\frac{1}{2}S[\sigma,\varphi]\,.
\end{equation}
Plugging in the Gauss decomposition \eqref{eq:Gauss main}, the WZNW action \eqref{eq:WZNW} reduces to \cite{Barnich:2013yka}
\begin{equation}
\label{eq:free field}
S_{\CS}[A]=-\frac{k}{4\pi} \int dt d\varphi\, \left(\phi'\, \bar \partial \phi-4 e^{-\phi} \sigma'\, \bar \partial \tau \right)\,.
\end{equation}
The boundary conditions on $\sigma$ given in \eqref{eq:bc Liouville} further reduce the second term to a total derivative, such that \eqref{eq:free field} is the action of a massless chiral field $\phi$. The chiral fields $\phi$ and $\bar \phi$ can be combined into a single Liouville field $\phi_L$ through a Backlund transformation whose details can be found in \cite{Henneaux:1999ib,Barnich:2013yka}, resulting in
\begin{equation}
\label{eq:flat Liouville}
S_{\CS}[A]-S_{\CS}[\bA]=-\frac{k}{2\pi} \int dtd\varphi \left(\frac{1}{2}\partial \phi_L \bar \partial \phi_L+2 e^{\phi_L}\right)\,.
\end{equation}  
This is the famous result of the Hamiltonian reduction of three-dimensional gravity to flat Liouville theory with potential term, first obtained in \cite{Coussaert:1995zp}. 

The main difference between the boundary  Polyakov action of the preceding sections and the flat Liouville action \eqref{eq:flat Liouville} is that the latter is still off-shell. Indeed, classical bulk solutions satisfy the additional condition
\begin{equation}
a_t=g^{-1} \partial_t g=\sqrt{2}\left(\dot \tau- \tau \dot \phi-e^{-\phi} \tau^2  \dot \sigma\right) \Jt_0 + \sqrt{2} e^{-\phi} \dot \sigma \Jt_1-\left( \dot \phi+2 e^{-\phi}\tau \dot \sigma\right)\Jt_2\,,
\end{equation}
compatible with \eqref{eq:at aphi} if and only if
\begin{equation}
\bar \partial \sigma=\bar \partial \phi=\bar \partial \tau=0\,.
\end{equation}
In that case, the boundary actions \eqref{eq:free field} and \eqref{eq:flat Liouville} simply vanish, in agreement with the findings of the previous sections that the gravitational onshell action vanishes for flat boundary geometries. In addition, the free function $L$ reduces to the Schwarzian derivative of a `holomorphic' function $\sigma(z)$, which is indeed the appropriate expression for the expectation value of the chiral component of a CFT stress tensor in a flat background geometry,
\begin{equation}
L(z)=-\frac{1}{2} S[\sigma,z]\,.
\end{equation}
	
\subsection{Another Hamiltonian reduction}
\label{section:Cotler}
In this subsection I would like to briefly discuss a variant of the Hamiltonian reduction due to Cotler and Jensen \cite{Cotler:2018zff}, and clarify its relation with the other approaches discussed previously. This will set the basis for the ideas to be developed in the discussion that heavily rely on the results presented in \cite{Cotler:2018zff}.

The setup is the same as that of section~\ref{section:Hamiltonian reduction}. In particular, the action considered is again \eqref{eq:S hamiltonian} with boundary conditions
\begin{equation}
\label{eq:bc Cotler}
A=
\begin{pmatrix}
\frac{dr}{2r}+O(r^{-2}) & O(r^{-1})\\
r dz+O(r^{-1})\, & -\frac{dr}{2r}+O(r^{-2})
\end{pmatrix}\,, \quad  \bA=
\begin{pmatrix}
-\frac{dr}{2r}+O(r^{-2})\, & r d\bz+O(r^{-1})\\
O(r^{-1}) & \frac{dr}{2r}+O(r^{-2})
\end{pmatrix}\,,
\end{equation}
that satisfy the ansatz \eqref{eq:a restricted}. Adopting the same setup they however come to a different result, namely \cite{Cotler:2018zff}
\begin{equation}
\label{eq:result Cotler}
S_{\CS}[A]-S_{\CS}[\bA]=\frac{k}{4\pi} \int dtd\varphi\, \left(\frac{\phi'' \bar \partial \phi'}{(\phi')^2}-\phi' \bar \partial \phi\right)-\frac{k}{4\pi} \int dtd\varphi\, \left(\frac{\bar \phi'' \partial \bar \phi'}{(\bar \phi')^2}-\bar \phi'  \partial \bar \phi\right)\,.
\end{equation}
This expression looks very similar to the chiral generating functional $W[\mu]$ on the cylinder \eqref{eq:W mu phi}, and one might suspect $\phi$ to parametrize a nontrivial boundary geometry. We will see that an interpretation of this sort is indeed possible.

Similarly to the Hamiltonian reduction described in section~\ref{section:Hamiltonian reduction}, the constraint $F_{r\varphi}=0$ is solved by writing
\begin{equation}
A_r=G^{-1} \partial_r G\,, \qquad A_\varphi=G^{-1} \partial_\varphi G\,.
\end{equation}
A Gauss parametrization of the group elements $G, \bar G$ is employed,
\begin{equation}
G=
	\begin{pmatrix}
		1 & 0\\
		F & 1
	\end{pmatrix} 
	\begin{pmatrix}
		\lambda & 0\\
		0 & \lambda^{-1}
	\end{pmatrix}
	\begin{pmatrix}
		1 & \Psi\\
		0 & 1
	\end{pmatrix}\,, \qquad 	\bG=
	\begin{pmatrix}
	1 & -\bF\\
	0 & 1
	\end{pmatrix} 
	\begin{pmatrix}
	\blambda^{-1} & 0\\
	0 & \blambda
	\end{pmatrix}
	\begin{pmatrix}
	1 & 0\\
	-\bPsi & 1
	\end{pmatrix}\,,
\end{equation}
such that the gauge connections take the form
\begin{align}
	A&=\begin{pmatrix}
		d \ln \lambda- \Psi \lambda^2 dF\quad & 2 \Psi d \ln \lambda+d \Psi-\Psi^2 \lambda^2 dF\\
		\lambda^2 dF & -d \ln \lambda+ \Psi \lambda^2 dF
	\end{pmatrix}\,,\\
	\bA&=\begin{pmatrix}
-d \ln \blambda+ \bPsi \blambda^2 d\bF & -\blambda^2 d\bF \\
-2 \bPsi d \ln \blambda-d \bPsi+\bPsi^2 \blambda^2 d\bF \quad & d \ln \blambda- \bPsi \blambda^2 d\bF
\end{pmatrix}\,.
\end{align}
Strictly speaking, at this point the above expressions only account for the $r, \varphi$ components since $A_t=G^{-1} \partial_t G$ does not necessarily hold. Imposing the boundary conditions \eqref{eq:bc Cotler} on the two spatial components implies 
\begin{align}
F=O(r^{0})\,, \qquad \lambda^2=\frac{r}{F'}+O(r^{-1})\,, \qquad \Psi=-\frac{F''}{2rF'}+O(r^{-2})\,,
\end{align}
and similarly for the barred quantities. Making the change of variable $F=\tan \phi$ and plugging this back into the WZNW expression \eqref{eq:WZNW} yields the result \eqref{eq:result Cotler}. 

Just as the flat Liouville action \eqref{eq:flat Liouville}, the boundary action \eqref{eq:result Cotler} is off-shell with respect to the bulk equations of motion. Bulk solutions do satisfy the additional conditions
\begin{subequations}
\begin{align}
A_t&=G^{-1} \partial_t G=
	\begin{pmatrix}
		O(r^{0}) & O(r^{-1})\\
		r \frac{\dot{\phi}}{\phi'}+O(r^0) & O(r^0)
	\end{pmatrix}\,,\\
\bA_t&=\bar{G}^{-1} \partial_t \bar G=
	\begin{pmatrix}
		O(r^0) & -r \frac{\dot{\bar{\phi}}}{\bar \phi'}+O(r^0)\\
		O(r^{-1}) & O(r^0)
	\end{pmatrix}\,.
\end{align}
\end{subequations}
We conclude that bulk solutions satisfies the boundary conditions \eqref{eq:bc Cotler} if and only if 
\begin{equation}
\label{eq:condition phi}
\bar \partial \phi=0\,, \qquad \partial \bar \phi=0\,,
\end{equation}
in which case the action \eqref{eq:result Cotler} again identically vanishes. However, if we do not impose \eqref{eq:condition phi} right away and collect all the components of the onshell connections $A$ and $\bA$, using \eqref{eq:A definitions} we find the expression for the associated bulk tetrad
\begin{equation}
e=\frac{r}{\sqrt{2}}\left(d\varphi+\frac{\dot{\bar \phi}}{\bar \phi'}dt\right)\Jt_0+\frac{r}{\sqrt{2}}\left(d\varphi+\frac{\dot{\phi}}{\phi'}dt\right)\Jt_1+O(r^0)\,,
\end{equation}
such that the boundary metric reads
\begin{subequations}
\begin{align}
	d\tilde{s}^2&=\left(d\varphi+\frac{\dot{\phi}}{\phi'}\, dt\right)\left(d\varphi+\frac{\dot{\bar \phi}}{\bar \phi'}\, dt\right)\\
	&=\frac{\partial \phi}{\partial \phi-\bar \partial \phi}\frac{\bar \partial \bar \phi}{\bar \partial \bar \phi-\partial \bar \phi}\left(dz+\frac{\bar \partial \phi}{\partial \phi}d\bz \right)\left(d\bz+\frac{\partial \bar \phi}{\bar \partial \bar \phi} dz \right)\,.
\end{align}
\end{subequations}
This should strongly remind us of the parametrization of a curved geometry \eqref{eq:metric z zbar} in terms of Beltrami differentials, although the conformal factor is not arbitrary.

This suggests an alternative way of recovering the boundary action \eqref{eq:result Cotler}, based on the holographic treatment described in section~\ref{section:metric} or section~\ref{section:Banados} and the resulting Polyakov action. Indeed, as discussed in section~\ref{section:Polyakov}, the local Quillen--Belavin--Knizhnik anomaly can be subtracted from the generating functional $W[\tilde g_{ij}]$ in order to achieve holomorphic factorization at the expense of diffeomorphism invariance,
\begin{equation}
	\label{eq:W holo}
	W_{\text{holo}}=W[\mu]+W[\bar \mu]\,.
\end{equation}
Recalling the parametrization appropriate to the cylinder \eqref{eq:W mu phi}, we have
\begin{equation}
\label{eq:alternative}
W[\mu]=-\frac{c}{24\pi} \int d^2w \left(\frac{\partial \bar \partial \phi\, \partial^2 \phi}{(\partial \phi)^2}-\bar \partial \phi\, \partial \phi \right)\,,
\end{equation}
and the result of Cotler and Jensen \eqref{eq:result Cotler} is recovered upon replacement $\partial_\varphi \mapsto \partial$ in the first term and $\partial_\varphi \mapsto \bar \partial$ in the second term.\footnote{I believe that this slight modification would not affect the other results in \cite{Cotler:2018zff}.} Note that this alternative derivation of the (cylinder) Alekseev--Shatashvili action is radically different from the one due to Cotler and Jensen and reviewed above. Indeed, in the latter the field $\phi$ describes off-shell dynamical bulk modes when a flat boundary metric is assumed, while $\phi$ in \eqref{eq:alternative} should be interpreted as parametrizing a curved boundary metric playing the role of background source for the dual CFT stress tensor. This alternative derivation simply follows from the classical GKPW dictionary \eqref{GKPW classical} that has been the basis for most investigations within the AdS/CFT correspondence. It also suggests another interpretation for those computations performed by Cotler and Jensen that are based on the Alekseev--Shatashvili action \eqref{eq:result Cotler}. I come back to this point in the discussion. 

\section{Discussion}
\label{section:discussion}

The generating functional of stress tensor correlation functions is an important and universal object characterizing any two-dimensional CFT, and in this review I have presented a unified view regarding its holographic derivation within the AdS$_3$/CFT$_2$ correspondence in both metric and Chern--Simons formulations. The literature on this subject is vast and confusing, and with the present work I hope to have given a robust account that will allow further developments. Below I discuss a few open problems which appear relevant to recent developments in holography.
 
\paragraph{Multiple boundaries and wormholes.} I restricted this review to the case of a single asymptotic boundary with the topology of the plane or cylinder. The case of a single torus boundary is straightforward to obtain and yields the same expression \eqref{eq:W mu} for the chiral generating functional $W[\mu]$ with the integration domain restricted to a single fundamental domain of torus \cite{Aldrovandi:1996sa}. Generalizations of the AdS/CFT generating functional to multiple asymptotic boundary components connected through spacetime wormholes are yet missing, although Hamiltonian reductions have been generalized to that context \cite{Henneaux:2019sjx,Cotler:2020ugk}. An important difference compared to the case of a single boundary is the presence of nontrivial holonomies in the Chern--Simons connections which eventually couple the boundary zero modes of the disconnected boundary components \cite{Henneaux:1999ib,Henneaux:2019sjx,Cotler:2020ugk}. Although they have been the basis of recent discussions about the role of wormholes in quantum gravity \cite{Cotler:2020ugk,Cotler:2021cqa}, it should be emphasized that Hamiltonian reductions have no straightforward interpretation within the AdS/CFT correspondence. It might therefore be interesting to work out the generating functional associated with multiple boundaries and wormholes, and subsequently use it as the basis for further holographic studies in that context.

\paragraph{Ensemble average in AdS$_3$/CFT$_2$.} Recently there has been a lot of interest in a new kind of holographic duality between gravity with AdS asymptotics and dual \textit{ensembles} of strongly coupled quantum systems, where the prime example is a correspondence between Jackiw--Teitelboim gravity in AdS$_2$ and the Sachdev--Ye--Kitaev quantum mechanichal ensemble \cite{Maldacena:2016hyu,Maldacena:2016upp,Engelsoy:2016xyb,Jensen:2016pah,Kitaev:2017awl,Sarosi:2017ykf}. There is increasing evidence that quantum mechanical ensembles are \textit{not} exact duals of quantum gravitational theories but rather describe a form of coarse-graining from which statistical properties can be obtained \cite{Stanford:2020wkf,Belin:2020hea,Belin:2020jxr,Altland:2020ccq,Altland:2021rqn,Saad:2021rcu}. This should remind us of the work of Wigner who showed that the energy level spacing statistics of heavy nuclei can be obtained from random matrix theory \cite{Wigner}, although there is no doubt that the fundamental description of nuclei does not involve any ensemble of theories. Similarly, it has been conjectured long ago that the spectral statistics of any chaotic quantum mechanical system are described by random matrix ensembles \cite{Berry,Bohigas:1983er}. Quantum gravity being chaotic, it is exponentially difficult to access detailed information about pure sates that would depart from the coarse-grained description. 

In the context of AdS$_3$ gravity, Cotler and Jensen have given a path integral derivation of the spectral statistics of black holes, where the action used is a modification of \eqref{eq:result Cotler} appropriate to the case of two disconnected asymptotic boundary components \cite{Cotler:2020ugk}. They found agreement with the predictions of a particular random matrix ensemble, leading them to the conclusion that AdS$_3$ gravity might be dual to an ensemble of CFTs rather than a single one. This conclusion, which is in tension with the common lore on the AdS/CFT correspondence \cite{Maldacena:1997re}, may have been premature. First of all, the preceding discussion shows that ensemble averaging \textit{can} often be used to obtain spectral statistics of quantum chaotic systems that are otherwise fundamentally described by a single theory. In addition, I argued at the end of section~\ref{section:Cotler} that the boundary action \eqref{eq:result Cotler} coincides with the Polyakov action in disguise, such that it alternatively follows from the standard GKPW dictionary \eqref{GKPW classical} which would give $\phi$ the interpretation of a \textit{source} rather than a dynamical field. As reviewed in section~\ref{section:Polyakov}, in that context $\phi$ is related to the source $\mu$ on the plane through equations \eqref{eq:Beltrami}, \eqref{eq:F definition} and \eqref{eq:phi definition}. It is therefore very tempting to re-interpret the computation of Cotler and Jensen as an \textit{explicit averaging} of the holographic CFT generating functional $Z_{\text{CFT}}[\mu]=e^{-W[\mu]}$ over the source~$\mu$, with integral measure \cite{Cotler:2020ugk}
\begin{equation}
\label{eq:measure}
\frac{d\phi}{\partial \phi}=df\,,
\end{equation}
where $f$ is related to $\mu$ through the Beltrami equation \eqref{eq:Beltrami}. This contrasts with the logic of these authors in which $\phi$ plays the role of fluctuating bulk field while the boundary geometry is kept flat. However, to make the above re-interpretation fully precise requires some more work. In particular, one should properly discuss the generating functional associated with two disconnected boundaries (see previous paragraph). I hope to report on this problem in a future publication.

As an additional remark, recall that $W[\mu]$ is also the action for the two-dimensional quantum gravity of Polyakov in the lightcone gauge\footnote{For a description of Polyakov gravity in conformal gauge, see \cite{Distler:1988jt,Seiberg:1990eb,DHoker:1990prw,Ginsparg:1993is}.}, with \eqref{eq:measure} the appropriate path integral measure \cite{Polyakov:1987zb,Polyakov:1988qz,Alekseev:1988ce,Knizhnik:1988ak}. Therefore, the computation of Cotler and Jensen is also a computation in 2d quantum gravity. It would be extremely interesting to revisit some of the old results in the latter theory in light of these new developments. Perhaps unexpectedly, we might learn about 3d quantum gravity from 2d quantum gravity.

\paragraph{Finite central charge.} The derivation of the Polyakov generating functional within the AdS/CFT correspondence has been presented in the limit of large central charge $c \to \infty$, i.e., in the classical gravity regime. However, the form of the generating functional of any two-dimensional CFT is the same whether at large or finite central charge, such that there cannot be any $O(c^{-1})$ correction to this result. In some sense the classical saddle point approximation \eqref{GKPW classical} appears exact in AdS$_3$/CFT$_2$. This can be viewed as a very stringent constraint to be satisfied by any nonperturbative definition of quantum gravity in AdS$_3$, i.e., by the right-hand side of \eqref{GKPW}. Such a nonperturbative definition is crucially missing at this time, which prevents any real progress towards a detailed understanding of quantum gravity within the AdS/CFT correspondence. 

\paragraph{Higher dimensions.} Conformal anomalies exist in all even dimensions. Like in two-dimensions, they can be integrated into nonlocal effective actions \cite{Mazur:2001aa}. In contrast to two dimensions, stress tensor correlation functions are not fully determined by the anomaly coefficients and therefore their most general form cannot be generated from the nonlocal actions. However, the latter encode most of the information about low-point functions. In four dimensions for example, they partially determine up to three-point correlators of the stress tensor  \cite{Osborn:1993cr,Coriano:2017mux}. It would be interesting to repeat the holographic derivation reviewed in section~\ref{section:metric} in higher dimensions in order to understand their emergence within the AdS/CFT correspondence. Related discussions can be found in \cite{Mazur:2001aa,Manvelyan:2001pv}.

\paragraph{Holography beyond AdS.} Rather interestingly, the effective action $W[\mu]$ appears at the boundary of spacetimes that are not asymptotically AdS. In particular, it appears at the spacelike boundaries of three-dimensional asymptotically de Sitter gravity \cite{Cotler:2019nbi} and on the celestial sphere in four-dimensional asymptotically flat gravity \cite{Nguyen:2020hot}. Since this effective action is characteristic of two-dimensional CFTs, it strongly hints at the holographic nature of these gravitational theories. 

\acknowledgments{I thank Teresa Bautista, Jordan Cotler, Chris Herzog, Jakob Salzer and Gideon Vos for useful discussions. This work is supported by a grant from the Science and Technology Facilities Council (STFC).}

\appendix	

\section{Chiral generating functional and diffeomorphism anomaly}
\label{app:Yoshida}
Following Yoshida \cite{Yoshida:1988xm}, I review the derivation of the chiral generating functional $W[\mu]$ from the conformal Ward identity satisfied by the stress tensor correlation functions themselves. The very definition for the chiral generating functional is that it generates all the correlation functions of the chiral component of the stress tensor,
\begin{gather}
\label{def:generating functional}
e^{-W[\mu]}\equiv \sum_{n=0}^\infty \frac{(-\pi)^{-n}}{n!} \int d^2z_1...\,d^2z_n\, \mu(z_1,\bz_1)...\mu(z_n,\bz_n) \langle T(z_1)...T(z_n) \rangle\,.
\end{gather}

Stress tensor correlation functions can be determined recursively from the conformal Ward identities \cite{Belavin:1984vu},
\begin{align}
	\label{eq:conf Ward}
	\langle T(z)T(z_1)...T(z_n) \rangle&=\sum_{i=1}^n \frac{c}{2(z-z_i)^4}\, \langle T(z_1)...\cancel{T(z_i)}...T(z_n) \rangle\\
	\nonumber
	&+\sum_{i=1}^n \left(\frac{2}{(z-z_i)^2}+\frac{\partial_{z_i}}{z-z_i}\right) \langle T(z_1)...T(z_n)  \rangle\,,
\end{align}
whose differentiation directly yields the anomalous chiral diffeomorphism Ward identity
\begin{align}
	\label{eq:diff Ward}
	\bar \partial \langle T(z)T(z_1)...T(z_n) \rangle=&-\sum_{i=1}^n \frac{\pi c}{12}\, \partial^3 \delta(x-x_i) \, \langle T(z_1)...\cancel{T(z_i)}...T(z_n) \rangle\\
	\nonumber
	&-\pi \sum_{i=1}^n \left(2\, \partial \delta(z-z_i)-\delta(z-z_i)\partial_{z_i}\right) \langle T(z_1)...T(z_n) \rangle\,,
\end{align}
where I made used of the distributional identity $\bar \partial (1/z)=\pi \delta(x)$. In turn, this directly translates to a statement for the the generating functional \eqref{def:generating functional},
\begin{align}
\label{eq:Ward identity}
\left(\bar \partial- \mu \partial-2 \partial \mu \right) \frac{\delta W}{\delta \mu(z,\bz)}=\frac{c}{12\pi}\, \partial^3 \mu\,.
\end{align}
Using the chain rule
\begin{align}
\frac{\delta}{\delta f(z,\bz)}&=\int d^2w\, \frac{\delta \mu(w,\bw)}{\delta f(z,\bz)} \frac{\delta}{\delta \mu(w,\bw)}=-\frac{1}{\partial f}\left[ \bar \partial -\mu \partial -2\partial \mu\right] \frac{\delta}{\delta \mu(z,\bz)}\,,
\end{align}
the Ward identity \eqref{eq:Ward identity} can also be written in terms of the quasiconformal mapping $f$ defined in \eqref{eq:Beltrami}, 
\begin{equation}
\frac{\delta W}{\delta f(z,\bz)}=-\frac{c}{12\pi}\, \partial f\, \partial^3 \mu\,.
\end{equation}
As can be explicitly checked, the solution to this equation is
\begin{equation}
\label{eq:Psi no puncture}
W\left[\mu\right]=\frac{c}{24\pi} \int d^2z\, \frac{\bar \partial f}{\partial f}\, \partial^2 \ln \partial f\,.
\end{equation}

We can also explicitly compute the variation of $W[\mu]$ \cite{Lazzarini:1990xid}. For that, we first note that \eqref{eq:Beltrami} implies
\begin{equation}
(\bar \partial -\mu \partial)\ln \partial f=\partial \mu\,,
\end{equation}
and thus
\begin{equation}
(\bar \partial -\mu \partial) \delta \ln \partial f=\partial \delta \mu+\delta \mu \partial \ln \partial f\,.
\end{equation}
Using these two relations and a few integration by parts, we obtain
\begin{align}
\delta W[\mu]&=\frac{c}{12 \pi} \int d^2z\, \delta \mu \left(\partial^2 \ln \partial f-\frac{1}{2}(\partial \ln \partial f)^2\right)=\frac{c}{12 \pi} \int d^2z\, \delta \mu\, S[f,z]\,,
\end{align}
where $S[f,z]$ is the Schwarzian derivative \eqref{eq:T vev main}. Hence, we obtain the familiar result for the expectation value of the stress tensor in a background geometry $\mu$ related to the complex plane by a (quasi)conformal mapping $f$,
\begin{equation}
\label{eq:T vev}
\langle T \rangle_{\mu}=\pi\, \frac{\delta W[\mu]}{\delta \mu}=\frac{c}{12}\, S[f,z]\,.
\end{equation} 

\section{Active vs.~passive diffeomorphisms}
\label{app:diffeos}
I briefly review the distinction and relation between diffeomorphisms and coordinate transformations (passive diffeomorphisms) which is often a source of confusion. This plays a role in section~\ref{section:metric} where the freedom in the conformal mode of the boundary metric associated with PBH diffeomorphisms \eqref{eq:PBH} is traded for a freedom in the coordinate location of the cutoff boundary surface \eqref{eq:boundary location}. This distinction between active and passive diffeomorphisms, along with its prominent role in Einstein's struggle to make physical sense of General Relativity, is beautifully discussed in \cite{Rovelli:2004tv}. 

\paragraph{Coordinate transformations.} Under an invertible change of coordinates
\begin{equation}
x \mapsto x'(x)\,,
\end{equation}
the components of a contravariant tensor field $A$ of rank $p$ evaluated at a point $P \in \mathcal{M}$ are related in the two charts by
\begin{equation}
\label{eq:coord transf}
{A'}\indices{_{\alpha_1...\alpha_p}}(P)=\left[\frac{\partial x^{\mu_1}}{\partial x'^{\alpha_1}}\, ...\, \frac{\partial x^{\mu_p}}{\partial x'^{\alpha_p}}\, A\indices{_{\mu_1...\mu_p}}\right](P)\,.
\end{equation}
This simply follows from the chain rule of differential calculus. Both sides of this equation are evaluated at the same point $P$, with coordinates $x(P)$ and $x'(P)$ in the respective charts. In the case of a scalar field $\Phi$ for example, this equation simply reads
\begin{equation}
\Phi'(P)=\Phi(P)\,.
\end{equation} 
The numerical value of $\Phi'$ and $\Phi$ at the point $P$ is the same, although the value the coordinates $x(P)$ and $x'(P)$ are different.

\paragraph{Maps between manifolds.} Before getting to diffeomorphisms themselves, we discuss generic smooth maps $\phi : \mathcal{M} \to \mathcal{N}$ between two manifolds. Given a function $f: \mathcal{N} \to \mathbb{R}$, the pullback function $\phi^* f:\mathcal{M} \to \mathbb{R}$ is defined by
\begin{equation}
\phi^* f=f \circ \phi\,.
\end{equation}
In particular, integration of $\phi^* f$ over a domain $\mathcal{D} \subset \mathcal{M}$ yields
\begin{equation}
\label{eq:integral diff}
\int_{\mathcal{D}} \phi^* f=\int_{\phi(\mathcal{D})} f\,.
\end{equation}
Given a vector field $V \in T(\mathcal{M})$, the pushforward vector field $\phi_* V \in T(\mathcal{N})$ is defined by 
\begin{equation}
\phi_* V (f)=V(\phi^* f)\,, \qquad \forall\, f: \mathcal{N} \to \mathbb{R}\,. 
\end{equation}
Finally, given a contravariant tensor field $A \in T^{*p}(\mathcal{N})$, the pullback tensor field $\phi^* A$ over $\mathcal{M}$ is defined by
\begin{equation}
\phi^* A(V_1,...,V_p)=A(\phi_* V_1,...,\phi_* V_p)\,, \qquad \forall\, V_1,...,V_p \in T(\mathcal{M})\,.
\end{equation}
There has been no need to introduce coordinate systems in order to build these definitions. However, if we chart $\mathcal{M}$ and $\mathcal{N}$ with coordinates $y^\alpha$ and $x^\mu$, respectively, in components the above equation becomes
\begin{equation}
\label{eq:map M to N}
(\phi^* A)_{\alpha_1...\alpha_p}(P)=\left[\frac{\partial x^{\mu_1}}{\partial y^{\alpha_1}}\, ...\, \frac{\partial x^{\mu_p}}{\partial y^{\alpha_p}}\, A\indices{_{\mu_1...\mu_p}}\right](\phi(P))\,, \qquad \forall\, P \in \mathcal{M}\,.
\end{equation}
This equation looks dangerously similar to the transformation law of $A$ under a change of coordinates \eqref{eq:coord transf}, although they should be clearly distinguished. Indeed, the left and right-hand sides of \eqref{eq:map M to N} are evaluated at \textit{different points}, which in fact belong to distinct manifolds.

\paragraph{Diffeomorphisms.} In the case that $\mathcal{N}=\mathcal{M}$, the smooth map $\phi$ discussed above is called a diffeomorphism if it is invertible. Considering a point $P \in \mathcal{M}$ with coordinates $x$, and designating by $x'$ another set of coordinates defined by
\begin{equation}
x'=\phi(x)\,,
\end{equation}
the relation \eqref{eq:map M to N} between $A$ and $\phi^* A$ becomes
\begin{equation}
\label{eq:diffeo}
A'_{\alpha_1...\alpha_p}(P)\equiv (\phi^* A)_{\alpha_1...\alpha_p}(P)=\left[\frac{\partial x^{\mu_1}}{\partial x'^{\alpha_1}}\, ...\, \frac{\partial x^{\mu_p}}{\partial x'^{\alpha_p}}\, A\indices{_{\mu_1...\mu_p}}\right](\phi(P))\,, \quad \forall\, P \in \mathcal{M}\,.
\end{equation}
This looks exactly like the transformation law \eqref{eq:coord transf} of the components of $A$ under a change of coordinates $x \mapsto x'(x)$ except that the right-hand side of \eqref{eq:diffeo} is evaluated at $\phi(P)$ rather than $P$ itself. Diffeomorphisms are not mere changes of coordinates, they actually drag tensor fields along with them. However, their action can be and is often represented by coordinate changes, keeping in mind this additional subtlety. 

In section~\ref{section:metric}, the freedom in the conformal mode of the boundary metric $\tilde g_{ij}$ that is associated with PBH diffeomorphisms \eqref{eq:PBH} is traded for a freedom in the location of the cutoff boundary surface $\partial \mathcal{M}$ by straightforward application of the integral identity \eqref{eq:integral diff}.

\bibliography{bibl}

\providecommand{\href}[2]{#2}\begingroup\raggedright\begin{thebibliography}{10}

\bibitem{Verlinde:1989ua}
H.L.~Verlinde, \emph{{Conformal Field Theory, 2-$D$ Quantum Gravity and
  Quantization of Teichmuller Space}},
  \href{https://doi.org/10.1016/0550-3213(90)90510-K}{\emph{Nucl. Phys. B}
  {\bfseries 337} (1990) 652}.

\bibitem{Coussaert:1995zp}
O.~Coussaert, M.~Henneaux and P.~van Driel, \emph{{The Asymptotic dynamics of
  three-dimensional Einstein gravity with a negative cosmological constant}},
  \href{https://doi.org/10.1088/0264-9381/12/12/012}{\emph{Class.Quant.Grav.}
  {\bfseries 12} (1995) 2961}
  [\href{https://arxiv.org/abs/gr-qc/9506019}{{\ttfamily gr-qc/9506019}}].

\bibitem{Maldacena:1997re}
J.M.~Maldacena, \emph{{The Large N limit of superconformal field theories and
  supergravity}}, {\emph{Adv.Theor.Math.Phys.} {\bfseries 2} (1998) 231}
  [\href{https://arxiv.org/abs/hep-th/9711200}{{\ttfamily hep-th/9711200}}].

\bibitem{Gubser:1998bc}
S.~Gubser, I.R.~Klebanov and A.M.~Polyakov, \emph{{Gauge theory correlators
  from noncritical string theory}},
  \href{https://doi.org/10.1016/S0370-2693(98)00377-3}{\emph{Phys.Lett.}
  {\bfseries B428} (1998) 105}
  [\href{https://arxiv.org/abs/hep-th/9802109}{{\ttfamily hep-th/9802109}}].

\bibitem{Witten:1998qj}
E.~Witten, \emph{{Anti-de Sitter space and holography}},
  {\emph{Adv.Theor.Math.Phys.} {\bfseries 2} (1998) 253}
  [\href{https://arxiv.org/abs/hep-th/9802150}{{\ttfamily hep-th/9802150}}].

\bibitem{Carlip:2005zn}
S.~Carlip, \emph{{Conformal field theory, (2+1)-dimensional gravity, and the
  BTZ black hole}},
  \href{https://doi.org/10.1088/0264-9381/22/12/R01}{\emph{Class. Quant. Grav.}
  {\bfseries 22} (2005) R85}
  [\href{https://arxiv.org/abs/gr-qc/0503022}{{\ttfamily gr-qc/0503022}}].

\bibitem{Henneaux:1999ib}
M.~Henneaux, L.~Maoz and A.~Schwimmer, \emph{{Asymptotic dynamics and
  asymptotic symmetries of three-dimensional extended AdS supergravity}},
  \href{https://doi.org/10.1006/aphy.2000.5994}{\emph{Annals Phys.} {\bfseries
  282} (2000) 31} [\href{https://arxiv.org/abs/hep-th/9910013}{{\ttfamily
  hep-th/9910013}}].

\bibitem{Rooman:1999km}
M.~Rooman and P.~Spindel, \emph{{Aspects of (2+1)-dimensional gravity: AdS(3)
  asymptotic dynamics in the framework of Fefferman-Graham-Lee theorems}},
  {\emph{Annalen Phys.} {\bfseries 9} (2000) 161}
  [\href{https://arxiv.org/abs/hep-th/9911142}{{\ttfamily hep-th/9911142}}].

\bibitem{Rooman:2000zi}
M.~Rooman and P.~Spindel, \emph{{Holonomies, anomalies and the Fefferman-Graham
  ambiguity in AdS(3) gravity}},
  \href{https://doi.org/10.1016/S0550-3213(00)00636-2}{\emph{Nucl. Phys. B}
  {\bfseries 594} (2001) 329}
  [\href{https://arxiv.org/abs/hep-th/0008147}{{\ttfamily hep-th/0008147}}].

\bibitem{Barnich:2013yka}
G.~Barnich and H.A.~Gonzalez, \emph{{Dual dynamics of three dimensional
  asymptotically flat Einstein gravity at null infinity}},
  \href{https://doi.org/10.1007/JHEP05(2013)016}{\emph{JHEP} {\bfseries 05}
  (2013) 016} [\href{https://arxiv.org/abs/1303.1075}{{\ttfamily 1303.1075}}].

\bibitem{Cotler:2018zff}
J.~Cotler and K.~Jensen, \emph{{A theory of reparameterizations for AdS$_3$
  gravity}}, \href{https://doi.org/10.1007/JHEP02(2019)079}{\emph{JHEP}
  {\bfseries 02} (2019) 079}
  [\href{https://arxiv.org/abs/1808.03263}{{\ttfamily 1808.03263}}].

\bibitem{Henneaux:2019sjx}
M.~Henneaux, W.~Merbis and A.~Ranjbar, \emph{{Asymptotic dynamics of AdS$_3$
  gravity with two asymptotic regions}},
  \href{https://doi.org/10.1007/JHEP03(2020)064}{\emph{JHEP} {\bfseries 03}
  (2020) 064} [\href{https://arxiv.org/abs/1912.09465}{{\ttfamily
  1912.09465}}].

\bibitem{Brown:1986nw}
J.D.~Brown and M.~Henneaux, \emph{{Central Charges in the Canonical Realization
  of Asymptotic Symmetries: An Example from Three-Dimensional Gravity}},
  \href{https://doi.org/10.1007/BF01211590}{\emph{Commun.Math.Phys.} {\bfseries
  104} (1986) 207}.

\bibitem{Polyakov:1981rd}
A.M.~Polyakov, \emph{{Quantum Geometry of Bosonic Strings}},
  \href{https://doi.org/10.1016/0370-2693(81)90743-7}{\emph{Phys. Lett. B}
  {\bfseries 103} (1981) 207}.

\bibitem{Skenderis:1999nb}
K.~Skenderis and S.N.~Solodukhin, \emph{{Quantum effective action from the AdS
  / CFT correspondence}},
  \href{https://doi.org/10.1016/S0370-2693(99)01467-7}{\emph{Phys. Lett. B}
  {\bfseries 472} (2000) 316}
  [\href{https://arxiv.org/abs/hep-th/9910023}{{\ttfamily hep-th/9910023}}].

\bibitem{Krasnov:2000zq}
K.~Krasnov, \emph{{Holography and Riemann surfaces}},
  \href{https://doi.org/10.4310/ATMP.2000.v4.n4.a5}{\emph{Adv. Theor. Math.
  Phys.} {\bfseries 4} (2000) 929}
  [\href{https://arxiv.org/abs/hep-th/0005106}{{\ttfamily hep-th/0005106}}].

\bibitem{Krasnov:2001cu}
K.~Krasnov, \emph{{On holomorphic factorization in asymptotically AdS 3-D
  gravity}}, \href{https://doi.org/10.1088/0264-9381/20/18/311}{\emph{Class.
  Quant. Grav.} {\bfseries 20} (2003) 4015}
  [\href{https://arxiv.org/abs/hep-th/0109198}{{\ttfamily hep-th/0109198}}].

\bibitem{Manvelyan:2001pv}
R.~Manvelyan, R.~Mkrtchian and H.~Muller-Kirsten, \emph{{Holographic trace
  anomaly and cocycle of Weyl group}},
  \href{https://doi.org/10.1016/S0370-2693(01)00550-0}{\emph{Phys. Lett. B}
  {\bfseries 509} (2001) 143}
  [\href{https://arxiv.org/abs/hep-th/0103082}{{\ttfamily hep-th/0103082}}].

\bibitem{Banados:2002ey}
M.~Banados, O.~Chandia and A.~Ritz, \emph{{Holography and the Polyakov
  action}}, \href{https://doi.org/10.1103/PhysRevD.65.126008}{\emph{Phys. Rev.
  D} {\bfseries 65} (2002) 126008}
  [\href{https://arxiv.org/abs/hep-th/0203021}{{\ttfamily hep-th/0203021}}].

\bibitem{Banados:2004nr}
M.~Banados and R.~Caro, \emph{{Holographic ward identities: Examples from 2+1
  gravity}}, \href{https://doi.org/10.1088/1126-6708/2004/12/036}{\emph{JHEP}
  {\bfseries 12} (2004) 036}
  [\href{https://arxiv.org/abs/hep-th/0411060}{{\ttfamily hep-th/0411060}}].

\bibitem{Carlip:2005tz}
S.~Carlip, \emph{{Dynamics of asymptotic diffeomorphisms in (2+1)-dimensional
  gravity}}, \href{https://doi.org/10.1088/0264-9381/22/14/014}{\emph{Class.
  Quant. Grav.} {\bfseries 22} (2005) 3055}
  [\href{https://arxiv.org/abs/gr-qc/0501033}{{\ttfamily gr-qc/0501033}}].

\bibitem{Cotler:2020ugk}
J.~Cotler and K.~Jensen, \emph{{AdS$_{3}$ gravity and random CFT}},
  \href{https://doi.org/10.1007/JHEP04(2021)033}{\emph{JHEP} {\bfseries 04}
  (2021) 033} [\href{https://arxiv.org/abs/2006.08648}{{\ttfamily
  2006.08648}}].

\bibitem{Aharony:1999ti}
O.~Aharony, S.S.~Gubser, J.M.~Maldacena, H.~Ooguri and Y.~Oz, \emph{{Large N
  field theories, string theory and gravity}},
  \href{https://doi.org/10.1016/S0370-1573(99)00083-6}{\emph{Phys.Rept.}
  {\bfseries 323} (2000) 183}
  [\href{https://arxiv.org/abs/hep-th/9905111}{{\ttfamily hep-th/9905111}}].

\bibitem{Belavin:1984vu}
A.A.~Belavin, A.M.~Polyakov and A.B.~Zamolodchikov, \emph{{Infinite Conformal
  Symmetry in Two-Dimensional Quantum Field Theory}},
  \href{https://doi.org/10.1016/0550-3213(84)90052-X}{\emph{Nucl. Phys.}
  {\bfseries B241} (1984) 333}.

\bibitem{Verlinde:1989hv}
H.L.~Verlinde and E.P.~Verlinde, \emph{{Conformal Field Theory and Geometric
  Quantization}},  in \emph{{Trieste School and Workshop on Superstrings}},
  pp.~422--449, 10, 1989.

\bibitem{Knecht:1990wb}
M.~Knecht, S.~Lazzarini and F.~Thuillier, \emph{{Shifting the Weyl anomaly to
  the chirally split diffeomorphism anomaly in two-dimensions}},
  \href{https://doi.org/10.1016/0370-2693(90)90936-Z}{\emph{Phys. Lett. B}
  {\bfseries 251} (1990) 279}.

\bibitem{Quillen1985DeterminantsOC}
D.~Quillen, \emph{Determinants of cauchy-riemann operators over a riemann
  surface}, {\emph{Functional Analysis and Its Applications} {\bfseries 19}
  (1985) 31}.

\bibitem{Belavin:1986cy}
A.A.~Belavin and V.G.~Knizhnik, \emph{{Algebraic Geometry and the Geometry of
  Quantum Strings}},
  \href{https://doi.org/10.1016/0370-2693(86)90963-9}{\emph{Phys. Lett. B}
  {\bfseries 168} (1986) 201}.

\bibitem{Polyakov:1987zb}
A.M.~Polyakov, \emph{{Quantum Gravity in Two-Dimensions}},
  \href{https://doi.org/10.1142/S0217732387001130}{\emph{Mod. Phys. Lett. A}
  {\bfseries 2} (1987) 893}.

\bibitem{Polyakov:1988qz}
A.M.~Polyakov, \emph{{Two-dimensional quantum gravity: Superconductivity at
  high T/c}},  in \emph{{Les Houches Summer School in Theoretical Physics:
  Fields, Strings, Critical Phenomena}}, 1988.

\bibitem{Alekseev:1988ce}
A.~Alekseev and S.L.~Shatashvili, \emph{{Path Integral Quantization of the
  Coadjoint Orbits of the Virasoro Group and 2D Gravity}},
  \href{https://doi.org/10.1016/0550-3213(89)90130-2}{\emph{Nucl. Phys.}
  {\bfseries B323} (1989) 719}.

\bibitem{Alvarez:1985ez}
O.~Alvarez, \emph{{Differential Geometry in String Models}},  in
  \emph{{Workshop on Unified String Theories}}, 10, 1985.

\bibitem{Nelson:1986ab}
P.C.~Nelson, \emph{{Lectures on Strings and Moduli Space}},
  \href{https://doi.org/10.1016/0370-1573(87)90082-2}{\emph{Phys. Rept.}
  {\bfseries 149} (1987) 337}.

\bibitem{Giddings:1987im}
S.B.~Giddings, \emph{{Conformal Techniques in String Theory and String Field
  Theory}}, \href{https://doi.org/10.1016/0370-1573(88)90096-8}{\emph{Phys.
  Rept.} {\bfseries 170} (1988) 167}.

\bibitem{DHoker:1988pdl}
E.~D'Hoker and D.H.~Phong, \emph{{The Geometry of String Perturbation Theory}},
  \href{https://doi.org/10.1103/RevModPhys.60.917}{\emph{Rev. Mod. Phys.}
  {\bfseries 60} (1988) 917}.

\bibitem{Nguyen:2020jqp}
K.~Nguyen, \emph{{Reparametrization modes in 2d CFT and the effective theory of
  stress tensor exchanges}},
  \href{https://doi.org/10.1007/JHEP05(2021)029}{\emph{JHEP} {\bfseries 21}
  (2020) 029} [\href{https://arxiv.org/abs/2101.08800}{{\ttfamily
  2101.08800}}].

\bibitem{Aldrovandi:1996sa}
E.~Aldrovandi and L.A.~Takhtajan, \emph{{Generating functional in CFT and
  effective action for two-dimensional quantum gravity on higher genus Riemann
  surfaces}}, \href{https://doi.org/10.1007/s002200050156}{\emph{Commun. Math.
  Phys.} {\bfseries 188} (1997) 29}
  [\href{https://arxiv.org/abs/hep-th/9606163}{{\ttfamily hep-th/9606163}}].

\bibitem{Balasubramanian:1999re}
V.~Balasubramanian and P.~Kraus, \emph{{A Stress tensor for Anti-de Sitter
  gravity}}, \href{https://doi.org/10.1007/s002200050764}{\emph{Commun. Math.
  Phys.} {\bfseries 208} (1999) 413}
  [\href{https://arxiv.org/abs/hep-th/9902121}{{\ttfamily hep-th/9902121}}].

\bibitem{deHaro:2000vlm}
S.~de~Haro, S.N.~Solodukhin and K.~Skenderis, \emph{{Holographic reconstruction
  of space-time and renormalization in the AdS / CFT correspondence}},
  \href{https://doi.org/10.1007/s002200100381}{\emph{Commun. Math. Phys.}
  {\bfseries 217} (2001) 595}
  [\href{https://arxiv.org/abs/hep-th/0002230}{{\ttfamily hep-th/0002230}}].

\bibitem{FeffermanGraham}
C.~Fefferman and C.~Graham, \emph{{Conformal Invariants}},  in \emph{{Elie
  Cartan et les Mathematiques d’Aujourd’hui}}, 1985.

\bibitem{Imbimbo:1999bj}
C.~Imbimbo, A.~Schwimmer, S.~Theisen and S.~Yankielowicz,
  \emph{{Diffeomorphisms and holographic anomalies}},
  \href{https://doi.org/10.1088/0264-9381/17/5/322}{\emph{Class. Quant. Grav.}
  {\bfseries 17} (2000) 1129}
  [\href{https://arxiv.org/abs/hep-th/9910267}{{\ttfamily hep-th/9910267}}].

\bibitem{Skenderis:2000in}
K.~Skenderis, \emph{{Asymptotically Anti-de Sitter space-times and their stress
  energy tensor}}, \href{https://doi.org/10.1142/S0217751X0100386X}{\emph{Int.
  J. Mod. Phys. A} {\bfseries 16} (2001) 740}
  [\href{https://arxiv.org/abs/hep-th/0010138}{{\ttfamily hep-th/0010138}}].

\bibitem{Troessaert:2013fma}
C.~Troessaert, \emph{{Enhanced asymptotic symmetry algebra of $AdS$$_{3}$}},
  \href{https://doi.org/10.1007/JHEP08(2013)044}{\emph{JHEP} {\bfseries 08}
  (2013) 044} [\href{https://arxiv.org/abs/1303.3296}{{\ttfamily 1303.3296}}].

\bibitem{Papadimitriou:2005ii}
I.~Papadimitriou and K.~Skenderis, \emph{{Thermodynamics of asymptotically
  locally AdS spacetimes}},
  \href{https://doi.org/10.1088/1126-6708/2005/08/004}{\emph{JHEP} {\bfseries
  0508} (2005) 004} [\href{https://arxiv.org/abs/hep-th/0505190}{{\ttfamily
  hep-th/0505190}}].

\bibitem{Alessio:2020ioh}
F.~Alessio, G.~Barnich, L.~Ciambelli, P.~Mao and R.~Ruzziconi, \emph{{Weyl
  charges in asymptotically locally AdS$_3$ spacetimes}},
  \href{https://doi.org/10.1103/PhysRevD.103.046003}{\emph{Phys. Rev. D}
  {\bfseries 103} (2021) 046003}
  [\href{https://arxiv.org/abs/2010.15452}{{\ttfamily 2010.15452}}].

\bibitem{Achucarro:1987vz}
A.~Achucarro and P.~Townsend, \emph{{A Chern-Simons Action for
  Three-Dimensional anti-De Sitter Supergravity Theories}},
  \href{https://doi.org/10.1016/0370-2693(86)90140-1}{\emph{Phys.Lett.}
  {\bfseries B180} (1986) 89}.

\bibitem{Witten:1988hc}
E.~Witten, \emph{{(2+1)-Dimensional Gravity as an Exactly Soluble System}},
  \href{https://doi.org/10.1016/0550-3213(88)90143-5}{\emph{Nucl. Phys.}
  {\bfseries B311} (1988) 46}.

\bibitem{Donnay:2016iyk}
L.~Donnay, \emph{{Asymptotic dynamics of three-dimensional gravity}},
  \href{https://doi.org/10.22323/1.271.0001}{\emph{PoS} {\bfseries Modave2015}
  (2016) 001} [\href{https://arxiv.org/abs/1602.09021}{{\ttfamily
  1602.09021}}].

\bibitem{Cotler:2021cqa}
J.~Cotler and K.~Jensen, \emph{{Wormholes and black hole microstates in
  AdS/CFT}},  \href{https://arxiv.org/abs/2104.00601}{{\ttfamily 2104.00601}}.

\bibitem{Maldacena:2016hyu}
J.~Maldacena and D.~Stanford, \emph{{Remarks on the Sachdev-Ye-Kitaev model}},
  \href{https://doi.org/10.1103/PhysRevD.94.106002}{\emph{Phys. Rev. D}
  {\bfseries 94} (2016) 106002}
  [\href{https://arxiv.org/abs/1604.07818}{{\ttfamily 1604.07818}}].

\bibitem{Maldacena:2016upp}
J.~Maldacena, D.~Stanford and Z.~Yang, \emph{{Conformal symmetry and its
  breaking in two dimensional Nearly Anti-de-Sitter space}},
  \href{https://doi.org/10.1093/ptep/ptw124}{\emph{PTEP} {\bfseries 2016}
  (2016) 12C104} [\href{https://arxiv.org/abs/1606.01857}{{\ttfamily
  1606.01857}}].

\bibitem{Engelsoy:2016xyb}
J.~Engels\"oy, T.G.~Mertens and H.~Verlinde, \emph{{An investigation of
  AdS$_{2}$ backreaction and holography}},
  \href{https://doi.org/10.1007/JHEP07(2016)139}{\emph{JHEP} {\bfseries 07}
  (2016) 139} [\href{https://arxiv.org/abs/1606.03438}{{\ttfamily
  1606.03438}}].

\bibitem{Jensen:2016pah}
K.~Jensen, \emph{{Chaos in AdS$_2$ Holography}},
  \href{https://doi.org/10.1103/PhysRevLett.117.111601}{\emph{Phys. Rev. Lett.}
  {\bfseries 117} (2016) 111601}
  [\href{https://arxiv.org/abs/1605.06098}{{\ttfamily 1605.06098}}].

\bibitem{Kitaev:2017awl}
A.~Kitaev and S.J.~Suh, \emph{{The soft mode in the Sachdev-Ye-Kitaev model and
  its gravity dual}},
  \href{https://doi.org/10.1007/JHEP05(2018)183}{\emph{JHEP} {\bfseries 05}
  (2018) 183} [\href{https://arxiv.org/abs/1711.08467}{{\ttfamily
  1711.08467}}].

\bibitem{Sarosi:2017ykf}
G.~S\'arosi, \emph{{AdS$_{2}$ holography and the SYK model}},
  \href{https://doi.org/10.22323/1.323.0001}{\emph{PoS} {\bfseries Modave2017}
  (2018) 001} [\href{https://arxiv.org/abs/1711.08482}{{\ttfamily
  1711.08482}}].

\bibitem{Stanford:2020wkf}
D.~Stanford, \emph{{More quantum noise from wormholes}},
  \href{https://arxiv.org/abs/2008.08570}{{\ttfamily 2008.08570}}.

\bibitem{Belin:2020hea}
A.~Belin and J.~de~Boer, \emph{{Random statistics of OPE coefficients and
  Euclidean wormholes}},
  \href{https://doi.org/10.1088/1361-6382/ac1082}{\emph{Class. Quant. Grav.}
  {\bfseries 38} (2021) 164001}
  [\href{https://arxiv.org/abs/2006.05499}{{\ttfamily 2006.05499}}].

\bibitem{Belin:2020jxr}
A.~Belin, J.~De~Boer, P.~Nayak and J.~Sonner, \emph{{Charged Eigenstate
  Thermalization, Euclidean Wormholes and Global Symmetries in Quantum
  Gravity}},  \href{https://arxiv.org/abs/2012.07875}{{\ttfamily 2012.07875}}.

\bibitem{Altland:2020ccq}
A.~Altland and J.~Sonner, \emph{{Late time physics of holographic quantum
  chaos}}, \href{https://doi.org/10.21468/SciPostPhys.11.2.034}{\emph{SciPost
  Phys.} {\bfseries 11} (2021) 034}
  [\href{https://arxiv.org/abs/2008.02271}{{\ttfamily 2008.02271}}].

\bibitem{Altland:2021rqn}
A.~Altland, D.~Bagrets, P.~Nayak, J.~Sonner and M.~Vielma, \emph{{From operator
  statistics to wormholes}},
  \href{https://doi.org/10.1103/PhysRevResearch.3.033259}{\emph{Phys. Rev.
  Res.} {\bfseries 3} (2021) 033259}
  [\href{https://arxiv.org/abs/2105.12129}{{\ttfamily 2105.12129}}].

\bibitem{Saad:2021rcu}
P.~Saad, S.H.~Shenker, D.~Stanford and S.~Yao, \emph{{Wormholes without
  averaging}},  \href{https://arxiv.org/abs/2103.16754}{{\ttfamily
  2103.16754}}.

\bibitem{Wigner}
E.P.~Wigner, \emph{Characteristics vectors of bordered matrices with infinite
  dimensions ii}, {\emph{Annals of Mathematics} {\bfseries 65} (1957) 203}.

\bibitem{Berry}
M.V.~Berry and M.~Tabor, \emph{Level clustering in the regular spectrum},
  {\emph{Proceedings of the Royal Society of London. Series A, Mathematical and
  Physical Sciences} {\bfseries 356} (1977) 375}.

\bibitem{Bohigas:1983er}
O.~Bohigas, M.J.~Giannoni and C.~Schmit, \emph{{Characterization of chaotic
  quantum spectra and universality of level fluctuation laws}},
  \href{https://doi.org/10.1103/PhysRevLett.52.1}{\emph{Phys. Rev. Lett.}
  {\bfseries 52} (1984) 1}.

\bibitem{Distler:1988jt}
J.~Distler and H.~Kawai, \emph{{Conformal Field Theory and 2D Quantum
  Gravity}}, \href{https://doi.org/10.1016/0550-3213(89)90354-4}{\emph{Nucl.
  Phys. B} {\bfseries 321} (1989) 509}.

\bibitem{Seiberg:1990eb}
N.~Seiberg, \emph{{Notes on quantum Liouville theory and quantum gravity}},
  \href{https://doi.org/10.1143/PTPS.102.319}{\emph{Prog. Theor. Phys. Suppl.}
  {\bfseries 102} (1990) 319}.

\bibitem{DHoker:1990prw}
E.~D'Hoker, \emph{{Equivalence of Liouville theory and 2-D quantum gravity}},
  \href{https://doi.org/10.1142/S0217732391000774}{\emph{Mod. Phys. Lett. A}
  {\bfseries 6} (1991) 745}.

\bibitem{Ginsparg:1993is}
P.H.~Ginsparg and G.W.~Moore, \emph{{Lectures on 2-D gravity and 2-D string
  theory}},  in \emph{{Theoretical Advanced Study Institute (TASI 92): From
  Black Holes and Strings to Particles}}, 10, 1993
  [\href{https://arxiv.org/abs/hep-th/9304011}{{\ttfamily hep-th/9304011}}].

\bibitem{Knizhnik:1988ak}
V.G.~Knizhnik, A.M.~Polyakov and A.B.~Zamolodchikov, \emph{{Fractal Structure
  of 2D Quantum Gravity}},
  \href{https://doi.org/10.1142/S0217732388000982}{\emph{Mod. Phys. Lett. A}
  {\bfseries 3} (1988) 819}.

\bibitem{Mazur:2001aa}
P.O.~Mazur and E.~Mottola, \emph{{Weyl cohomology and the effective action for
  conformal anomalies}},
  \href{https://doi.org/10.1103/PhysRevD.64.104022}{\emph{Phys. Rev. D}
  {\bfseries 64} (2001) 104022}
  [\href{https://arxiv.org/abs/hep-th/0106151}{{\ttfamily hep-th/0106151}}].

\bibitem{Osborn:1993cr}
H.~Osborn and A.C.~Petkou, \emph{{Implications of conformal invariance in field
  theories for general dimensions}},
  \href{https://doi.org/10.1006/aphy.1994.1045}{\emph{Annals Phys.} {\bfseries
  231} (1994) 311} [\href{https://arxiv.org/abs/hep-th/9307010}{{\ttfamily
  hep-th/9307010}}].

\bibitem{Coriano:2017mux}
C.~Coriano, M.M.~Maglio and E.~Mottola, \emph{{TTT in CFT: Trace Identities and
  the Conformal Anomaly Effective Action}},
  \href{https://doi.org/10.1016/j.nuclphysb.2019.03.019}{\emph{Nucl. Phys. B}
  {\bfseries 942} (2019) 303}
  [\href{https://arxiv.org/abs/1703.08860}{{\ttfamily 1703.08860}}].

\bibitem{Cotler:2019nbi}
J.~Cotler, K.~Jensen and A.~Maloney, \emph{{Low-dimensional de Sitter quantum
  gravity}}, \href{https://doi.org/10.1007/JHEP06(2020)048}{\emph{JHEP}
  {\bfseries 06} (2020) 048}
  [\href{https://arxiv.org/abs/1905.03780}{{\ttfamily 1905.03780}}].

\bibitem{Nguyen:2020hot}
K.~Nguyen and J.~Salzer, \emph{{The effective action of superrotation modes}},
  \href{https://doi.org/10.1007/JHEP02(2021)108}{\emph{JHEP} {\bfseries 02}
  (2021) 108} [\href{https://arxiv.org/abs/2008.03321}{{\ttfamily
  2008.03321}}].

\bibitem{Yoshida:1988xm}
K.~Yoshida, \emph{{Effective Action for Quantum Gravity in Two-dimensions}},
  \href{https://doi.org/10.1142/S0217732389000101}{\emph{Mod. Phys. Lett. A}
  {\bfseries 4} (1989) 71}.

\bibitem{Lazzarini:1990xid}
S.~Lazzarini, \emph{{Sur les Modeles Conformes Lagrangiens Bidimensionnels}},
  Ph.D. thesis, Savoie U., 1990.

\bibitem{Rovelli:2004tv}
C.~Rovelli, \emph{{Quantum gravity}}, Cambridge Monographs on Mathematical
  Physics, Univ. Pr., Cambridge, UK (2004).

\end{thebibliography}\endgroup
\bibliographystyle{JHEP}
\end{document}